\documentclass[preprint,12pt]{aastex}
\usepackage{natbib,amssymb}
\usepackage[usenames]{color}
\usepackage{lscape}

\def\simlt{\mathrel{\hbox{\rlap{\hbox{\lower4pt\hbox{$\sim$}}}\hbox{$<$}}}}
\def\simgt{\mathrel{\hbox{\rlap{\hbox{\lower4pt\hbox{$\sim$}}}\hbox{$>$}}}}
\newcommand\be{\begin{equation}}
\newcommand\ee{\end{equation}}
\newcommand{\mbf[1]}{\mathbf{#1}}

\begin{document}
\title{Precision Astrometry with Adaptive Optics}
\author{P. B. Cameron, M. C. Britton and S. R. Kulkarni}
\affil{California Institute of Technology, Division Physics,
Mathematics and Astronomy, MC 105-24, Pasadena, CA 91125 \\ Electronic
mail: pbc@astro.caltech.edu}

\begin{abstract}
We investigate the limits of ground-based astrometry with adaptive
optics using the core of the Galactic globular cluster M5.  Adaptive
optics systems provide near diffraction-limit imaging with the world's
largest telescopes. The substantial improvement in both resolution and
signal-to-noise ratio enables high-precision astrometry from the
ground.  We describe the dominant systematic errors that typically
limit ground-based differential astrometry, and enumerate
observational considerations for mitigating their effects.  After
implementing these measures, we find that the dominant limitation on
astrometric performance in this experiment is caused by tilt
anisoplanatism.  We then present an optimal estimation technique for
measuring the position of one star relative to a grid of reference
stars in the face of this correlated random noise source. Our
methodology has the advantage of reducing the astrometric errors as
$\sim 1/\sqrt{t}$ and faster than the square root of the number of
reference stars -- effectively eliminating noise caused by atmospheric
tilt to the point that astrometric performance is limited by centering
accuracy.  Using 50 reference stars we demonstrate single-epoch
astrometric precision of $\approx 1\,$mas in 1\,second, decreasing to
$\simlt 100\,\mu$as in 2\,minutes of integration time at the Hale
200-inch telescope.  We also show that our astrometry is accurate to
$\simlt 100\,\mu$as for observations separated by 2 months. Finally,
we discuss the limits and potential of differential astrometry with
current and next generation large aperture telescopes. At this level
of accuracy, numerous astrometric applications become accessible,
including planet detection, astrometric microlensing signatures, and
kinematics of distant Galactic stellar populations.
\end{abstract}

\section{Introduction}
The benefits of astrometry have long been clear to
astronomers. Measurements of parallax and proper motion yield model
independent determinations of fundamental quantities like distance and
velocity. It is not surprising that astrometry has motivated a wide
variety of observational programs using many different techniques to
answer fundamental questions in astrophysics.  Potential applications
span a wide range of physical scales including: planet detection,
reconstruction of the Milky Way's formation, and tests of $\Lambda$CDM
cosmology (e.g.  \citealt{ust+07}).

The most ubiquitous astrometric measurements have been carried out
with ground-based telescopes in the seeing limit.  \cite{mdv+92}
conducted visible light measurements of 72 stars ($V=15$--20) using
the 1.55\,m US Naval Observatory astrometric reflector.  This program
achieved single epoch measurement precision $\approx
4$\,milliarcseconds (mas), and parallax accuracies ranging from 0.5--3
mas over $\sim$~5 yr baselines.  \cite{ps96} performed visible light
measurements of stars in the cluster NGC 2420 ($V=13$--16) and
achieved single epoch precision of $\approx 150\,\mu$as in one hour,
which motivated an astrometric survey for low-mass companions to
nearby stars (e.g. \citealt{psh+04}).  More recently,
200--300\,$\mu$as astrometric precision has been demonstrated with
VLT/FORS in the visible \citep{l06,lmd+07}. Each of the above programs
employed relatively narrow-field visible imagers (a few square
arcminutes) to perform differential astrometry; however, the
increasing availability of wide angle imagers has motivated studies
over larger fields.  \citealt{abp+06} performed similar experiments
using a 33' $\times$ 34' visible camera on the ESO 2.2\,m telescope,
which resulted in 7\,mas single-epoch precision.

Ground-based interferometers provide an alternative method for
performing high precision astrometry, typically over very narrow
fields relative to a single reference star.  The Palomar Testbed
Interferometer has used phase-referencing to achieve astrometric
accuracies $\approx 100\,\mu$as for a 30\arcsec\ binary
\citep{lcb+00}, and $\approx 20\,\mu$as over years for binaries with
separations $\simlt 1$\arcsec\ \citep{mlk+06}.  Due to its 40 cm
apertures, this instrument is limited to targets with $K_s < 6$.
Large aperture, ground-based interferometers equipped with adaptive
optics systems, such as those at Keck \citep{cw03} and the VLT
\citep{gac+00}, can perform at similar levels to fainter limiting
magnitudes (e.g. \citealt{bts+07}).
 
Differential astrometric accuracies achieved in both single aperture
and interferometric ground-based programs are fundamentally limited by
atmospheric effects.  In the seeing limit, single aperture
observations suffer from image quality degradation and interferometers
lose visibility fringe coherence due to atmospheric turbulence.  In
addition, all ground-based programs suffer from systematic effects due
to differential atmospheric refraction and optical distortions.

Space-based observatories are one possible method for avoiding the
effects of atmospheric turbulence. {\it Hipparcos} was the first
space-based mission with astrometric goals, and achieved $\simlt
1$\,mas astrometry over the mission lifetime on bright targets ($V
\simlt 9$ mag; \citealt{plk+97}).  Currently, the only space-based
telescope that can perform high-precision astrometry is
\textit{Hubble}. Both the imagers and Fine Guidance Sensor have been
characterized and well-utilized for astrometry at the $\simlt 1$\,mas
level (e.g. \citealt{ak00,ak03b,baf+03}). Two complimentary future
space missions are aimed at achieving levels of astrometric
performance 2--3 orders of magnitude below the {\it Hubble}
performance levels.  GAIA will catalog roughly one billion stars to $V
\approx 20$\,mag over the entire sky with parallax accuracies ranging
from 10--$300\,\mu$as depending on the magnitude \citep{pdg+01}.  The
Space Interferometry Mission (SIM) will take a pointed approach, and
enable microarcsecond ($\mu$as) astrometry on Galactic and
extragalactic targets \citep{ust+07}.

Ground-based adaptive optics (AO) offer an alternative, more easily
accessible, and cost effective method for overcoming atmospheric
turbulence over small fields ($\simlt$ arcminute).  The current
generation of astronomical adaptive optics systems provide diffraction
limited image quality at near-infrared wavelengths. Achieving the
telescope's diffraction limit and the resulting boost in
signal-to-noise ratio prove to be a powerful combination for
astrometry. These two effects reduce the errors in determining stellar
centers, increase the number of possible reference stars at small
separations, and allow techniques for mitigating systematics (e.g. use
of narrow-band filters to eliminate chromatic refraction; see
\S\ref{sec:errors}).

The marked improvement in wavefront sensor technology and the
development of laser beacons has rapidly increased the usable sky
coverage of these systems (e.g. \citealt{wlb+06a}). The increase in
sky coverage, operation in the near-infrared, gain in signal-to-noise
ratio, and the diffraction-limited image quality make astrometry with
adaptive optics amenable to numerous Galactic applications spanning a
wide number of fields: detection of astrometric companions, the
improved determination of the mass-luminosity relation of stars, and
the formation and evolution of compact objects \citep{ust+07}.

Here we present an optimal estimation technique appropriate for
mitigating the astrometric errors arising in AO observations and
demonstrate its potential with multi-epoch imaging of the core of the
globular cluster M5 using the Hale 200-inch Telescope. We are able to
achieve $\simlt 100\,\mu$as astrometric precision in 2 minutes, and
have maintained this accuracy over 2 months. In \S\ref{sec:errors} we
discuss the dominant noise terms that arise in ground-based astrometry
and the experimental techniques we have adopted to control them. We
lay out the framework of our reduction model and illustrate its
salient properties with a numerical simulation in \S\ref{sec:grid}. We
describe the observations of M5 and the results of applying the
optimal estimation technique to the data in \S\ref{sec:obs} and
\S\ref{sec:res}. This is followed in \S\ref{sec:dis} by a discussion
of the role and potential of adaptive optics in ground-based
astrometry with current and future large aperture telescopes.

\section{Astrometric Error Terms in Ground Based Astrometry}
\label{sec:errors}
Ground-based optical and infrared imaging observations suffer from a
number of errors that limit the accuracy and precision of astrometric
measurements.  Relative to seeing-limited observations, the
diffraction-limited image quality afforded by adaptive optics modifies
the relative importance of these error terms.  This section describes
the four largest effects, and indicates observational considerations
utilized in this experiment aimed at mitigating them.

\subsection{Differential Tilt Jitter}
\label{sec:tj}
With AO, the image motion of the guide star is removed with a flat
tip-tilt mirror. This stabilizes the image of the guide star with
respect to the imager to high accuracy. Any residual tip-tilt error is
removed in subsequent analysis by calculating only differential
offsets between the target of astrometry (not necessarily the AO guide
star) and the reference stars. However, the difference in the tilt
component of turbulence along any two lines of sight in the field of
view causes a correlated, stochastic change in their measured
separation, known as differential atmospheric tilt jitter.

More specifically, in propagating through the atmosphere to reach the
telescope aperture, light from the target star and light from a
reference star at a finite angular offset traverse different columns
of atmospheric turbulence that are sheared.  Differential atmospheric
tilt jitter arises from the decorrelation in the tilt component of the
wavefront phase aberration arising from this shearing effect.  This
differential tilt leads to a random, achromatic, and anisotropic
fluctuation in the relative displacement of the two objects. The three
term approximation to the parallel and perpendicular components of the
variance arising from differential atmospheric tilt jitter, assuming
Kolmogorov turbulence, is given by \citep{s94}
\be {\sigma^2_{\parallel,{\rm TJ}} \brack \sigma^2_{\perp,{\rm TJ}}} = 2.67
\frac{\mu_2}{D^{1/3}}\left(\frac{\theta}{D}\right)^2{3 \brack 1} -
3.68 \frac{\mu_4}{D^{1/3}}\left(\frac{\theta}{D}\right)^4{5 \brack 1}
+ 2.35
\frac{\mu_{14/3}}{D^{1/3}}\left(\frac{\theta}{D}\right)^{14/3}{17/3
\brack 1}.
\label{eqn:tj}
\ee 
In this equation $D$ is the telescope diameter and $\theta$ is the
angular separation of the stars.  The turbulence moments $\mu_m$ are
defined as
\be 
\mu_m = \sec^{m+1}\xi \int_0^\infty dh C_n^2(h)h^m, 
\ee
where $h$ is the altitude, $\xi$ is the zenith angle, and
$C_{n}^{2}(h)$ is the vertical strength of atmospheric
turbulence. Typical $C_n^2(h)$ profiles yield $\sigma_{\parallel,{\rm
TJ}} \approx 20$--30\,mas for a 20\arcsec\ binary when observed with a
5\,m aperture. Note that the variance from differential tilt is a
random error, and thus is also $\propto \tau_{\rm TJ}/t$, where
$\tau_{\rm TJ}$ is the tilt jitter timescale (of order the wind
crossing time over the aperture; see \S\ref{sec:res}) and $t$ is the
integration time.

\subsection{Distortion}
The largest instrumental systematic that limits the accuracy of
astrometry in any optical system is geometric distortion.  These
distortions can be stable --- resulting from unavoidable errors in the
shape or placement of optics --- or dynamic --- resulting from the
flexure or replacement of optics.

If geometric distortions are stable, then a number of strategies can
be employed to mitigate their effect.  One method is to model the
distortion to high accuracy; the most notable example is the
calibration of \textit{HST} (e.g. \citealt{ak03a}). This is
particularly important for data sets obtained with multiple
instruments or those that use the technique of dithering, since
knowledge of the distortion is necessary to place stellar positions in
a globally correct reference frame. Alternatively, one could use a
consistent optical prescription from epoch-to-epoch by using the same
instrument and placing the field at the same location and orientation
on the detector. Here we use both a distortion solution and a single,
consistent dither position to achieve accurate astrometry.

Any changes in the geometric distortion must be tracked through
routine, consistent calibration.  The question of stability is
particularly important at the Hale 200-inch, since the AO system and
the imaging camera (PHARO; see \S\ref{sec:obs}) are mounted at the
Cassegrain focus, and PHARO undergoes a few warming/cooling cycles per
month (see \S\ref{sec:obs}). The PHARO distortion
solution\footnote{see also
http://www.astro.ucla.edu/$\sim$metchev/ao.html} by \cite{m06}
accounts for changes in the orientation of the telescope (which are
relatively small for our experimental design), but the overall
stability of the system is best verified with on-sky data. One of the
purposes of the data presented here is to track the system
stability. We find that the combination of the Hale Telescope, PALAO,
and PHARO is capable of delivering $\simlt 100\,\mu$as astrometry.

\subsection{Atmospheric Refraction}
Refraction by the Earth's atmosphere causes an angular deflection of
light from a star, resulting in an apparent change in its position.
The magnitude of this deflection depends on the wavelength and the
atmospheric column depth encountered by an incoming ray. The former
effect is chromatic, while the latter is achromatic. The error induced
by differential chromatic refraction (DCR) has proven to be an
important, and sometimes the dominant, astrometric limitation in
ground-based efforts (e.g. \citealt{mdv+92,ps96,abp+06,l06}). These
studies have shown DCR can contribute $\approx 0.1$--1\,mas of error
depending on the wavelength and strategy of the observations.

The observations presented here were conducted using a Br-$\gamma$
filter at 2.166\,$\mu$m with a narrow bandpass of 0.02\,$\mu$m to
suppress differential chromatic refraction.  The increased
signal-to-noise ratio provided by adaptive optics allows sufficient
reference stars to be detected even through such a narrow filter in a
short exposure time.  We reach $K_s \approx 15$\,magnitude in our
1.4\,s exposures through this filter with the Hale 200-inch (see
\S\ref{sec:obs}). In addition, observations were acquired over a
relatively narrow range of airmass (1.17--1.27) at each epoch to
minimize the achromatic differential refraction.

In order to estimate the effect of atmospheric refraction on our data
we took the asterism in the core of M5 and refracted it to 37 and 32
degrees elevation with the parallactic angles appropriate for the
observations on 2007 May 28 using the {\it slarefro} function
distributed with the STARLINK library \citep{gt98}. The
root-mean-square (RMS) deviation in reference star positions between
these two zenith angles was $\simlt 250\,\mu$as and the shift in guide
star position with respect to the grid (see \S\ref{sec:grid}) was
$\approx 10\,\mu$as. Thus, our consistent zenith angle of
observations, narrow-band filter and observations in the near-infrared
(where the refraction is more benign) make the contribution of this
effect negligible for our purposes, and we make no effort to correct
for it.

Performing a similar experiment using a the broadband $K$ filter with
a field of $\approx 5000$\,K reference stars and a $\approx 3000$\,K
target would lead to a systematic shift of $\approx 100\,\mu$as
between zenith angles separated by 10$^\circ$, which would be
detectable by this experiment. Consequently, for observations where
broadband filters are necessary, refraction effects must be considered
and corrected.

\subsection{Measurement Noise}
\label{sec:mn}
In the case of a perfect optical system, a perfect detector and no
atmosphere, the astrometric precision is limited one's ability to
calculate stellar centers. The centering precision is determined by
measurement noise, and we will use the two terms interchangeably. For a
monopupil telescope the uncertainty is
\be 
\sigma_{\rm  meas} = \frac{\lambda}{\pi D}\frac{1}{\rm SNR} 
= 284\,\mu{\rm  as}\left(\frac{\lambda}{2.17\,\mu{\rm m}}\right) 
\left(\frac{5\,{\rm m}}{D}\right)\left(\frac{100}{\rm SNR}\right)
\label{eqn:mn}
\ee 
\citep{l78}.
Adaptive optics allow us to achieve the diffraction limit even in the
presence of the atmosphere and substantially boosts the SNR over the
seeing-limited case --- thereby decreasing measurement noise and
improving astrometric precision.

In practice, the centering of a given stellar image is limited by
spatial and temporal variations in the AO point-spread function (PSF).
A great deal of time and effort has been spent determining the AO PSF
and producing software packages to perform PSF fitting
(e.g. \citealt{dbb+00,b06a}). However, any PSF-fitting software
package is capable of calculating image positions at $\simlt 0.01$
pixel level in a single image. For the observations considered here
this is $\simlt 2$\,mas, a factor of 5 -- 10 larger than the
measurement noise in Equation~\ref{eqn:mn}, but it is much smaller
than the tilt jitter mentioned in \S\ref{sec:tj}. As such, we have
chosen to use simple and widely available PSF centering software
(DAOPHOT; \citealt{s87}; see \S\ref{sec:obs}).

\begin{deluxetable}{ccccccccccccccccc}
\tabletypesize{\scriptsize}
\tablewidth{0pt}
\tablecaption{Observations\label{tab:obs}}
\tablehead{
& &  \colhead{Integration Time} & & \colhead{Seeing\tablenotemark{a}} 
& \colhead{$\theta_{0}$\tablenotemark{a}} & \colhead{$\mu_{2}$} & \colhead{$\mu_{4}$} & 
\colhead{$\mu_{14/3}$} & \\
\colhead{ Date} & \colhead{ Time} & \colhead{(sec)} &\colhead{ Airmass} &  \colhead{(asecs)} & 
\colhead{(asecs)} & \colhead{($m^{7/3}$)} & \colhead{($m^{13/3}$)} & \colhead{($m^{15/3}$)} 
}
\startdata
2007-05-28 & 05:18:29 - 06:19:29 & 890 & 1.26 - 1.18  & 1.22    & 2.34    & 1.01e-5 & 3.82e3  & 2.77e6  \\
2007-05-29 & 05:58:26 - 06:48:30 & 570 & 1.19 - 1.17  & 1.39    & 2.16    & 1.14e-5 & 3.89e3  & 2.82e6  \\
2007-07-22 & 03:57:12 - 04:40:36 & 630 & 1.20 - 1.27  & 1.05    & 1.66    & 1.74e-5 & 6.46e3  & 4.71e6  \\
\enddata 
\tablenotetext{a}{Calculated at a wavelength of 0.5\,$\mu$m. These
quantities scale as $\lambda^{1/5}$ and $\lambda^{6/5}$, respectively.}
% wave=2.12 seeing=0.911 isoplanatic=13.2 isokinetic=20.4
\end{deluxetable}

\begin{figure}
\plotone{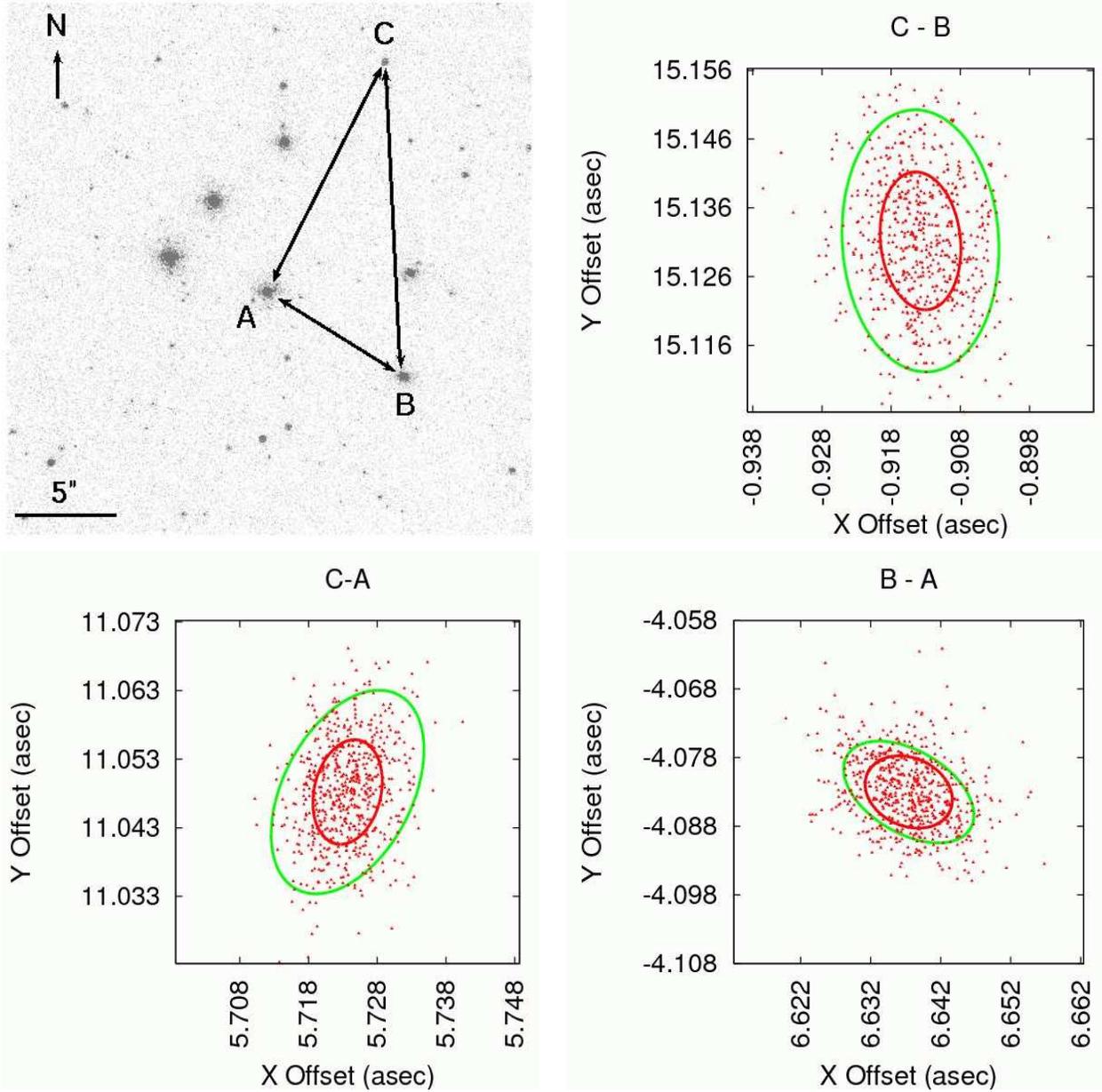}
\caption{{\it Top left:} Image of the core of the globular cluster M5
in 1.4 seconds through the narrow-band Br-$\gamma$ filter. The AO
guide star is labeled as star `A', and is one of 82 detected stars in
the image. The additional plots show the measured $x$-$y$ angular
separation of each pair of stars denoted by the arrows in 600 images
taken on 28 May 2007. These plots show the clear signature of
anisotropic differential atmospheric tilt jitter as predicted from
Equation~\ref{eqn:tj}. The measured (red) and predicted (green)
1\,$\sigma$ error ellipses are over-plotted. We see that temporal
averaging over the 1.4\,second exposure time has reduced the measured
variance with respect to that predicted from the DIMM/MASS
measurements and Equation~\ref{eqn:tj}.}
\label{fig:finder}
\end{figure}

\section{Observations and Data Reduction}
\label{sec:obs}
We observed the the globular cluster M5 on three dates spanning 2
months (see Table~\ref{tab:obs} for a summary of observations) using
the Hale 200-inch telescope and the Palomar High Angular Resolution
Observer (PHARO; \citealt{hbp+01}) assisted by the Palomar Adaptive
Optics System (PALAO; \citealt{tdb+00}).  The globular cluster M5 was
chosen for its relatively large distance of $\approx 7.5$\,kpc from
the Sun, low velocity dispersion of $\approx 5$\,km\,s$^{-1}$, and the
availability of guide stars near the cluster core
\citep{pm93,h96}. This combination of distance and velocity yields an
expected cluster dispersion of only $140\,\mu$as\,yr$^{-1}$, or
20\,$\mu$as over our 2 month observing span.  We acquired 400--600
images per night. A typical image can be found in
Figure~\ref{fig:finder}.  The guide star is a red giant branch member
of the globular cluster with $V \approx 12.6$ magnitude \citep{sb04}.
The cluster was imaged through the narrow-band Br-$\gamma$ filter
(central wavelength is 2.166\,$\mu$m and bandpass is 0.02\,$\mu$m)
using the 25\arcsec\ $\times$ 25\arcsec\ narrow-field channel
(0.025\arcsec\,pixel$^{-1}$), which over samples the 87\,mas
diffraction-limited PSF.  The brightest star filled the detector to
10\% of the maximum well-depth in the 1.4\,sec exposure time --- well
within the linear regime of the detector.

Contemporaneous measurements of the atmospheric turbulence profile
were acquired with a differential image motion monitor (DIMM) and
multi-aperture scintillation sensor (MASS), which have been deployed
as a single unit in a dome at Palomar Observatory
\citep{tbp07,kts+07}.  These turbulence profile measurements permitted
an independent estimate of the magnitude of differential tilt jitter
(computed using Equation~\ref{eqn:tj}).

We processed the raw images by subtracting dark frames and removing
bad pixels from the analysis.  Flat-field calibration was performed
using twilight sky flats. Sky subtraction was accomplished by forming
the median of the dithered frames taken outside of the cluster and
subtracting this median from each exposure.  The photometry and
astrometry of each star was extracted using PSF-fitting as implemented
by the DAOPHOT package in PyRAF\footnote{PyRAF is a product of Space
Telescope Science Institute, which is operated by AURA for
NASA.}. DAOPHOT is not optimized for astrometry (see e.g.
\citealt{ak00}), and since our measurement model reduces the noise due
to atmospheric turbulence, our single epoch precision could be
improved with a more careful centering technique (see \S\ref{sec:sim}
and \S\ref{sec:prec}). However, our astrometric accuracy over 2 months
is not limited by this choice (see \S\ref{sec:acc}). We used the 4
brightest stars in the field to derive a model PSF that is assumed to
be constant over the field, and calibrated the image zeropoints using
2MASS and find that the guide star has $K_s \approx 9.1$\,magnitude.

\section{Grid Astrometry for Ground-based Adaptive Optics Observations}
\label{sec:grid}
After controlling for distortion and atmospheric refraction, the
dominant astrometric errors are caused by differential atmospheric
tilt jitter and measurement noise. In this section we present a
general framework for measuring the position of a star relative to a
grid of reference stars in the face of these noise sources.  This
framework has two key ingredients. The first is the covariance matrix
($\mbf[\Sigma_d]$), which encapsulates the relevant statistical
uncertainties for astrometry with adaptive optics.  The second is the
weight matrix ($\mbf[W]$), which determines how the differential
measurements between the target star and the reference stars are
combined to calculate the target's position relative to the grid.

\subsection{Measurement Model}
The fundamental quantity in differential astrometry is the measured
angular offset between a pair of stars. We will denote the angular
distance between two stars, $i$ and $j$, as $\vec{d}_{ij}$. Since
$\vec{d}_{ij}$ is measured from an image, we will denote its
components in the Cartesian coordinate system of the detector, simply
\be
\vec{d}_{ij} = { x_j - x_i \brack y_j - y_i } \equiv { x_{ij} \brack y_{ij} },
\ee
where we have introduced the notation $x_{ij} \equiv x_j - x_i$ and
likewise for $y$. The variance in the the angular separation between
two stars is given by
\be 
{\sigma_{\parallel}^2 \brack \sigma_{\perp}^2} = 
{\sigma_{\parallel,{\rm meas}}^2 \brack \sigma_{\perp,\rm meas}^2} + 
\frac{\tau_{\rm TJ}}{t}
{\sigma_{\parallel,{\rm TJ}}^2 \brack \sigma_{\perp,{\rm TJ}}^2}
\label{eqn:varpair}
\ee 
where $\sigma_{\parallel,{\rm meas}}^2$ is the sum of the squares of
the centering errors of each star parallel to the axis connecting the
pair (and similarly for the perpendicular variance), and the remaining
terms are as defined in \S\ref{sec:tj}.

Measurement of the offset between the target star (which we will
denote with a subscript $i=0$) and each of the $N$ reference stars
results in a set of $N$ vectors, $\vec{d}_{0i}$. For simplicity, we
will write these measured offsets as a single column vector,
\be
\mbf[d] = [ x_{01},\cdots,x_{0N},y_{01},\cdots,y_{0N}]^{\rm T}.
\label{eqn:d}
\ee
The goal of differential astrometry is to use $\mbf[d]$ to determine
the position of the target star with respect to the reference grid
of stars at each epoch.

There are many possible ways to construct the position of the
astrometric target from a given $\mbf[d]$.  Here we use the most
general linear combination of the angular offsets, namely
\be 
\vec{p} = \mbf[Wd],
\label{eqn:wavg}
\ee
where $\mbf[W]$ is the 2 $\times 2N$ weight matrix, given by
\be
\mbf[W] = \left[\begin{array}{cccccc} 
w_{xx,01} & \cdots & w_{xx,0N} & w_{xy,01} & \cdots & w_{xy,0N} \\ 
w_{yx,01} & \cdots & w_{yx,0N} & w_{yy,01} & \cdots & w_{yy,0N}
\end{array} \right].
\label{eqn:W}
\ee
We have used the notation $w_{xy,0i}$ to denote the weighting of the
offset from the target star to star $i$ in the $y$ direction used to
determine the $x$ component of the target's position, $\vec{p}$.  For
example, for a standard average of the $x$ and $y$ measurements to
calculate $\vec{p}$, we would assign all the $w_{xx,0i} = w_{yy,0i} =
1/N$ and $w_{xy,0i} = w_{yx,0i} = 0$. 

In principle, we are free to assign weights in any manner we
please. However, we find it convenient to choose the weights such that
they satisfy
\be
\sum_i w_{xx,0i} = 1 \ \ , \ \ \sum_i w_{yy,0i} = 1 \ \ ,
\sum_i w_{xy,0i} = 0 \ \ , \ \ \sum_i w_{yx,0i} = 0.
\label{eqn:const}
\ee 
These constraints ensure that the components of $\vec{p}$ have
physical units (e.g. pixels or arcseconds) and that its components are
measured in the same coordinate system as $\mbf[d]$ (presumably the
detector coordinates). As a consequence, $\vec{p}$ represents the
position of the target star in the sense that a proper motion of the
target, $\vec{\epsilon}$, with respect to the fixed grid between two
epochs will cause a change, $\vec{p} \rightarrow \vec{p} +
\vec{\epsilon}$.

In order to determine if any change in $\vec{p}$ over time is
meaningful we must understand its statistical properties.  Both
differential tilt jitter and measurement errors are assumed to follow
Gaussian statistics, so that each instance of target-reference grid offset
measurements, $\mbf[d]$, is drawn from a multivariate normal
probability distribution:
\be 
P(\mbf[d]) =
\frac{1}{\sqrt{2\pi\det \mbf[\Sigma_d]}} \exp \left(\frac{1}{2}
[\mbf[d]-\bar{\mbf[d]}]^{\rm T} \mbf[\Sigma_d]^{-1}
[\mbf[d]-\bar{\mbf[d]}]\right),
\label{eqn:mnd}
\ee 
where $\mbf[\Sigma_d]$ is the covariance matrix, and the bars above
symbols denote using the average value of each matrix entry.

The statistics of $\vec{p}$ follow in a straightforward manner from
Equation~\ref{eqn:mnd} given our choice in Equation~\ref{eqn:wavg}.
Since $\vec{p}$ is a linear function of $\mbf[d]$, each $\vec{p}$ is
also drawn from a multivariate normal probability distribution
with covariance matrix
\be 
\mbf[\Sigma]_p = \mbf[W]^{\rm T}\mbf[\Sigma_d]\mbf[W],
\label{eqn:sigp}
\ee 
and the uncertainties of $\vec{p}$ are described by the eigenvectors
and eigenvalues of $\mbf[\Sigma]_p$. Thus, our goal of optimally
determining the target's position requires calculating the covariance
matrix, $\mbf[\Sigma_d]$, from data or theory, and choosing $\mbf[W]$
to minimize the eigenvalues of $\mbf[\Sigma]_p$.

\subsection{The Covariance Matrix}
\label{sec:cvm}
We have chosen to measure positions and offsets in the Cartesian
coordinates of the detector, so the form of the covariance
matrix, given our definitions above, is
{\small \be \mbf[\Sigma_d] = \left( \begin{array}{cccccc}
      \langle(\Delta x_{01})^2\rangle & \cdots &
      \langle(\Delta x_{01})(\Delta x_{0N})\rangle &
      \langle(\Delta x_{01})(\Delta y_{01})\rangle &
      \cdots &
      \langle(\Delta x_{01})(\Delta y_{0N})\rangle \\
      & \ddots & \vdots & \vdots & \ddots & \vdots \\
      & & \langle(\Delta x_{0N})^2\rangle & 
      \langle(\Delta x_{0N})(\Delta y_{01})\rangle & \cdots&
      \langle(\Delta x_{0N})(\Delta y_{0N})\rangle \\
      & & & \langle(\Delta y_{01})^2\rangle & \cdots &
      \langle(\Delta  y_{01})(\Delta y_{0N})\rangle\\
      & & & & \ddots & \vdots\\
      {\rm symmetric} & & & & & \langle(\Delta y_{0N})^2\rangle
\end{array} \right),
\label{eqn:cvm}
\ee } 
where we have written $\Delta x_{ij} \equiv (x_{ij} - \bar{x}_{ij})$
to simplify the notation (likewise for $y$).

The total covariance matrix has contributions from centering
errors and differential atmospheric tilt jitter. Since these
contributions are independent, the total covariance matrix can be
written $\mbf[\Sigma_d] = \mbf[\Sigma]_{\rm meas} + \mbf[\Sigma]_{\rm
TJ}$, and each term can be derived separately.

\subsubsection{The Covariance Matrix for Measurement Noise}
In the absence of differential tilt jitter it is straightforward to
construct the covariance matrix for measurement noise alone,
$\mbf[\Sigma]_{\rm meas}$. The diagonal terms can be written
\be
\langle\Delta x_{0i}^2\rangle \equiv \sigma_{x,0i}^2 = \sigma_{x,0}^2 +\sigma_{x,i}^2, 
\label{eqn:mdiag}
\ee
where $\sigma_{x,i}$ and $\sigma_{x,0}$ are the the
uncertainties in determining the $x$-position of star $i$ and the
target star, respectively. For the off-diagonal terms $\langle\Delta
x_{0i}\Delta x_{0j}\rangle$ we can use the fact that
\begin{eqnarray}
\langle\Delta x_{0i} \Delta x_{0j}\rangle 
&=& \frac{1}{2}\langle\{ \Delta x_{0i}^2 + \Delta x_{0j}^2 -
    [\Delta x_{0i}- \Delta x_{0j}]^2\}\rangle\cr
&=& \frac{1}{2}\{ \langle\Delta x_{0i}^2\rangle + \langle\Delta x_{0j}^2\rangle -
    \langle[\Delta x_{0i}- \Delta x_{0j}]^2\rangle\}\cr
&=& \frac{1}{2}\{ \langle\Delta x_{0i}^2\rangle + \langle\Delta x_{0j}^2\rangle -
    \langle\Delta x_{ij}^2\rangle\}\cr
&=& \frac{1}{2} (\sigma_{x,0i}^2 + \sigma_{x,0j}^2 - \sigma_{x,ij}^2)\cr
&=& \sigma_{x,0}^2.
\label{eqn:trick}
\end{eqnarray}
where we have used only algebra and the definitions
above. Equation~\ref{eqn:trick} is the obvious result of the fact that
the measurements of the target star's coordinates are common to all
differential measurements, and so its uncertainty appears in all the
off-diagonal covariance terms, $\langle\Delta x_{0i}\Delta
x_{0j}\rangle$ and $\langle\Delta y_{0i}\Delta
y_{0j}\rangle$. However, the cross-terms involving both $x$ and $y$
(e.g. $\langle\Delta x_{0i}\Delta y_{0j}\rangle$) vanish because
$\sigma_{x,0}$ and $\sigma_{y,0}$ are uncorrelated for measurement
noise alone.

\subsubsection{The Covariance Matrix for Differential Tilt Jitter}
The covariance matrix for differential atmospheric tilt
jitter between a pair of stars is diagonal when written in an
orthogonal coordinate system with one axis lying along the separation
axis of the binary.  We see from Equation~\ref{eqn:tj} that it can be
written as
\be \mbf[\Sigma]_{\rm pair} =
\left(\begin{array}{cc} \langle(d_\parallel -
\bar{d}_\parallel)^2\rangle & \langle(d_\parallel -
\bar{d}_\parallel)(d_\perp - \bar{d}_\perp)\rangle \\
\langle(d_\parallel - \bar{d}_\parallel)(d_\perp -
\bar{d}_\perp)\rangle & \langle(d_\perp - \bar{d}_\perp)^2\rangle
\end{array} \right) =
\left(\begin{array}{cc}
\sigma^2_{\parallel,{\rm TJ}} & 0 \\
0 & \sigma^2_{\perp,{\rm TJ}}
\end{array} \right),
\label{eqn:diag}
\ee 
where $d_\parallel$ and $d_\perp$ are the angular offsets parallel and
perpendicular to the axis connecting the pair of stars, respectively.

For a general field of $N$ stars, no coordinate system exists that
diagonalizes the full tilt jitter covariance matrix,
$\mbf[\Sigma]_{\rm TJ}$. But, we can begin computing the entries by
rotating $\mbf[\Sigma]_{\rm pair}$ into our $x$-$y$ coordinates via
$\mbf[R]^{\rm T}\mbf[\Sigma]_{\rm pair}\mbf[R]$, where
\be \mbf[R] =
\left(\begin{array}{cc} \cos\phi & \sin\phi\\ -\sin\phi & \cos\phi
\end{array} \right).
\ee
The result is
\be
\mbf[R]^{\rm T}\mbf[\Sigma]_{\rm pair}\mbf[R] = \left(\begin{array}{cc}
\sigma_{\parallel,0i}^2\cos^2\phi_{0i} + \sigma_{\perp,0i}^2\sin^2\phi_{0i} &
(\sigma_{\parallel,0i}^2 - \sigma_{\perp,0i}^2)\cos\phi_{0i}\sin\phi_{0i} \\
(\sigma_{\parallel,0i}^2 - \sigma_{\perp,0i}^2)\cos\phi_{0i}\sin\phi_{0i} &
\sigma_{\parallel,0i}^2\sin^2\phi_{0i} + \sigma_{\perp,0i}^2\cos^2\phi_{0i}
\end{array} \right).
\label{eqn:right}
\ee where $\phi_{0i}$ is the angle between $\vec{d}_{0i}$ and our
arbitrary Cartesian system measured counterclockwise from the
$x$-axis, and we have introduced the notation that the uncertainty
parallel to $\vec{d}_{ij}$ is $\sigma_{\parallel,ij}$ and uncertainty
orthogonal to $\vec{d}_{ij}$ is $\sigma_{\perp,ij}$ as calculated from
Equation~\ref{eqn:tj}. Thus, we can identify the diagonal terms
\begin{eqnarray}
\langle\Delta x_{0i}^2\rangle =
\sigma_{\parallel,0i}^2\cos^2\phi_{0i} + \sigma_{\perp,0i}^2\sin^2\phi_{0i},\cr
\langle\Delta y_{0i}^2\rangle =
\sigma_{\parallel,0i}^2\sin^2\phi_{0i} + \sigma_{\perp,0i}^2\cos^2\phi_{0i},
\label{eqn:dterms}
\end{eqnarray}
and
\be
\langle\Delta x_{0i} \Delta y_{0i}\rangle =
(\sigma_{\parallel,0i}^2 - \sigma_{\perp,0i}^2)\cos\phi_{0i}\sin\phi_{0i}.
\label{eqn:cterms}
\ee For the off-diagonal terms $\langle \Delta x_{0i} \Delta
x_{0j}\rangle$ we notice that (as used in
Equation~\ref{eqn:trick})
\begin{eqnarray}
\langle\Delta x_{0i} \Delta x_{0j}\rangle 
&=& \frac{1}{2}\langle\{ \Delta x_{0i}^2 + \Delta x_{0j}^2 -
    [\Delta x_{0i}- \Delta x_{0j}]^2\}\rangle\cr
&=& \frac{1}{2}\{ \langle\Delta x_{0i}^2\rangle + \langle\Delta x_{0j}^2\rangle -
    \langle[\Delta x_{0i}- \Delta x_{0j}]^2\rangle\}\cr
&=& \frac{1}{2}( \sigma_{\parallel,0i}^2\cos^2\phi_{0i} + 
    \sigma_{\perp,0i}^2\sin^2\phi_{0i} + \sigma_{\parallel,0j}^2\cos^2\phi_{0j} +
    \sigma_{\perp,0j}^2\sin^2\phi_{0j} \cr
& &  - \sigma_{\parallel,ij}^2\cos^2\phi_{ij} - \sigma_{\perp,ij}^2\sin^2\phi_{ij}),
\label{eqn:cterm2}
\end{eqnarray}
where in the last step we have used the fact that $x_{ij} =
x_{0i}-x_{0j}$ and the relations in Equations~\ref{eqn:dterms} and
\ref{eqn:cterms}. The quantities $\langle \Delta y_{0i} \Delta
y_{0j}\rangle$ can be obtained by interchanging sine and cosine in
Equation~\ref{eqn:cterm2}.

For the remaining off-diagonal terms $\langle\Delta x_{0i}\Delta
y_{0j}\rangle$ we can use the fact that
\begin{eqnarray}
\langle\Delta x_{0i}\Delta y_{0j}\rangle 
&=&\langle\Delta x_{0i}[\Delta y_{0i}+\Delta y_{ij}]\rangle\cr
&=&\langle\Delta x_{0i}\Delta y_{0i}\rangle+
   \langle\Delta x_{0i}\Delta y_{ij}]\rangle\cr
&=&\langle\Delta x_{0i}\Delta y_{0i}\rangle+
   \langle[\Delta x_{0j}-\Delta x_{ij}]\Delta y_{ij}\rangle\cr
&=&\langle\Delta x_{0i}\Delta y_{0i}\rangle+
   \langle\Delta x_{0j}\Delta y_{ij}\rangle-
   \langle\Delta x_{ij}\Delta y_{ij}\rangle\cr
&=&\langle\Delta x_{0i}\Delta y_{0i}\rangle+
   \langle\Delta x_{0j}[\Delta y_{0j}-\Delta y_{0i}]\rangle-
   \langle\Delta x_{ij}\Delta y_{ij}\rangle\cr
&=&\langle\Delta x_{0i}\Delta y_{0i}\rangle+
   \langle\Delta x_{0j}\Delta y_{0j}\rangle-
   \langle\Delta x_{ij}\Delta y_{ij}\rangle-
   \langle\Delta x_{0j}\Delta y_{0i}\rangle.
\end{eqnarray}
Rearranging gives
\begin{eqnarray}
\langle\Delta x_{0i}\Delta y_{0j}\rangle +
   \langle\Delta x_{0j}\Delta y_{0i}\rangle
&=&\langle\Delta x_{0i}\Delta y_{0i}\rangle+
   \langle\Delta x_{0j}\Delta y_{0j}\rangle-
   \langle\Delta x_{ij}\Delta y_{ij}\rangle.
\label{eqn:cterm3}
\end{eqnarray}
All the terms in the right-hand side are known from
Equation~\ref{eqn:cterms}, and further investigation shows that the
two terms on the left-hand side are equal.  So,
Equations~\ref{eqn:mdiag}, \ref{eqn:trick}, \ref{eqn:dterms},
\ref{eqn:cterms}, \ref{eqn:cterm2}, and \ref{eqn:cterm3} contain all
the information required to construct the full covariance matrix,
$\mbf[\Sigma_d]$.

\subsection{The Optimal Weight Matrix}
\label{sec:optimal}
The optimal choice of weights in Equation~\ref{eqn:mn} are those which
minimize the eigenvalues in Equation~\ref{eqn:sigp}.  For a $2 \times
2$ symmetric matrix the sum of the eigenvalues is the trace of the
matrix, so our problem becomes one of minimizing the trace of
$\mbf[\Sigma_p]$ subject to the constraints in
Equation~\ref{eqn:const}.  Specifically, we will use the method of
Lagrange multipliers \citep{b80} to find the optimal weights,
$\mbf[W]^\prime$, that minimize the quadratic equation
\be
{\rm Tr(}\mbf[\Sigma_p]) = \frac{1}{2} \mbf[W]^{\rm \prime T}\mbf[SW]^{\prime},
\label{eqn:min}
\ee
where
\be
\mbf[S] = \left[\begin{array}{cccc} 
\mbf[\Sigma_d] & \mbf[0] \\
\mbf[0]  & \mbf[\Sigma_d]
\end{array} \right],
\ee
is a $4N \times 4N$ matrix, and 
\be
\mbf[W]^\prime = [w_{xx,01},\cdots, w_{xx,0N}, \ w_{xy,01},\cdots,w_{xy,0N}, \ 
w_{yx,01},\cdots,w_{yx,0N}, \ w_{yy,01},\cdots,w_{yy,0N}]^{\rm T}
\ee
is a vector of length $4N$. Note that $\mbf[W]^\prime$ has identical
entries as $\mbf[W]$ in Equation~\ref{eqn:W}, it is just written as
a single vector to cast the minimization problem into a single
equation (\ref{eqn:min}). We want to find the extrema of
Equation~\ref{eqn:min} subject to the linear constraints in
Equation~\ref{eqn:const}, which can be written
\be
\mbf[CW]^\prime = \mbf[V],
\label{eqn:matrixconst}
\ee
where we define the $4 \times 4N$ symmetric matrix
\be
\mbf[C] = \left[\begin{array}{cccccccccccc} 
1 & \cdots & 1 & 0 & \cdots & 0 & 0 & \cdots & 0 & 0 & \cdots & 0\\
  &        &   & 1 & \cdots & 1 & 0 & \cdots & 0 & 0 & \cdots & 0\\
  &        &   &   &        &   & 1 & \cdots & 1 & 0 & \cdots & 0\\
{\rm sym}  &        &   &   &        &   &   &        &   & 1 & \cdots & 1
\end{array} \right],
\ee
and
\be
\mbf[V] = \left[\begin{array}{c} 
1 \\
0 \\
0 \\
1
\end{array} \right].
\ee

In this framework the optimal weights are those that solve the system
of linear equations
\be
\left[\begin{array}{cc} 
\mbf[S] & \mbf[C]^{\rm T}\\
\mbf[C] & \mbf[0]
\end{array} \right] 
\left[\begin{array}{c} 
\mbf[W]^\prime\\
\mbf[\lambda]
\end{array} \right] =
\left[\begin{array}{c} 
\mbf[0]\\
\mbf[V]
\end{array} \right].
\label{eqn:lm}
\ee
Here $\mbf[\lambda]$ are the Lagrange multipliers, which will not be
used further.  Equation~\ref{eqn:lm} can be solved via a matrix
inversion.

Note that the general constraints on the weights we have written in
Equation~\ref{eqn:const} and \ref{eqn:matrixconst} have two somewhat
unintuitive features.  The first is that the $y$ measurements are
sometimes used to compute the $x$ position and vice versa.  The other
property is that they allow for negative weights, meaning that in some
cases certain measurements will be subtracted in calculating the
position of the astrometric target, $\vec{p}$. These two facts conspire
to exploit the natural correlations inherent in the data. The flexible
and possibly negative weights essentially allow the reference grid to
be symmetrized, thereby using the known correlations to cancel noise
so as to minimize the variance in $\vec{p}$.

\subsection{Numerical Simulations}
\label{sec:sim}
As indicated in the above analysis, the single epoch uncertainty in
the location, $\vec{p}$, of the target relative to the grid of
reference stars is represented by the eigenvalues and eigenvectors of
the $2 \times 2$ matrix $\mbf[\Sigma]_p$ (Equation~\ref{eqn:sigp}).
This matrix itself depends on the distribution of reference stars, the
precision of centering measurements, and the degree of noise
correlation due to differential tilt through the matrix
$\mbf[\Sigma_d]$.  In this way, the intrinsic precision of the
measured value of $\vec{p}$ depends on these three factors.

To ascertain the behavior of $\mbf[\Sigma]_p$ with the density of
available reference stars, we performed a series of numerical
simulations.  In each simulation, $N$ ($2 \leq N \leq 100$) stars were
randomly distributed throughout a 25\arcsec\ $\times$ 25\arcsec\ field
of view. We assumed the target was a bright star in the middle of the
field with centering error of 0.5\,mas and the reference stars were
fainter, drawn from a Gaussian distribution with mean centering error
of 2\,mas and a standard deviation of 1\,mas (somewhat analogous to
the situation for the guide star in M5; see \S\ref{sec:res}).  The
full covariance matrix, $\mbf[\Sigma_d]$, was computed for each
stellar configuration assuming these centering errors, the typical
turbulence profile above Palomar Observatory, and a 1.4\,sec exposure
time.

In the first simulation, $\mbf[\Sigma_d]$ was contracted as in
Equation~\ref{eqn:sigp} using standard averaging for $\mbf[W]$
($w_{xx,0i} = w_{yy,0i} = 1/N$; $w_{xy,0i} = w_{yx,0i} = 0$). For the
second simulation, $\mbf[\Sigma_d]$ was contracted using the optimal
$\mbf[W]$ as calculated using the prescription in \S\ref{sec:optimal}.
In each case, the geometric mean of the two eigenvalues of the
resulting matrix, $\mbf[\Sigma]_p$, were computed to form an estimate
of the single epoch measurement precision of $\vec{p}$.  To average
away random effects arising from the particular geometry of the random
distribution of stars, each numerical simulation was repeated for 100
random distributions of stars for each value of $N$, and these were
averaged to generate a mean value for the single epoch measurement
precision.

\begin{figure}
\begin{center}
\includegraphics[angle=90,width=4in]{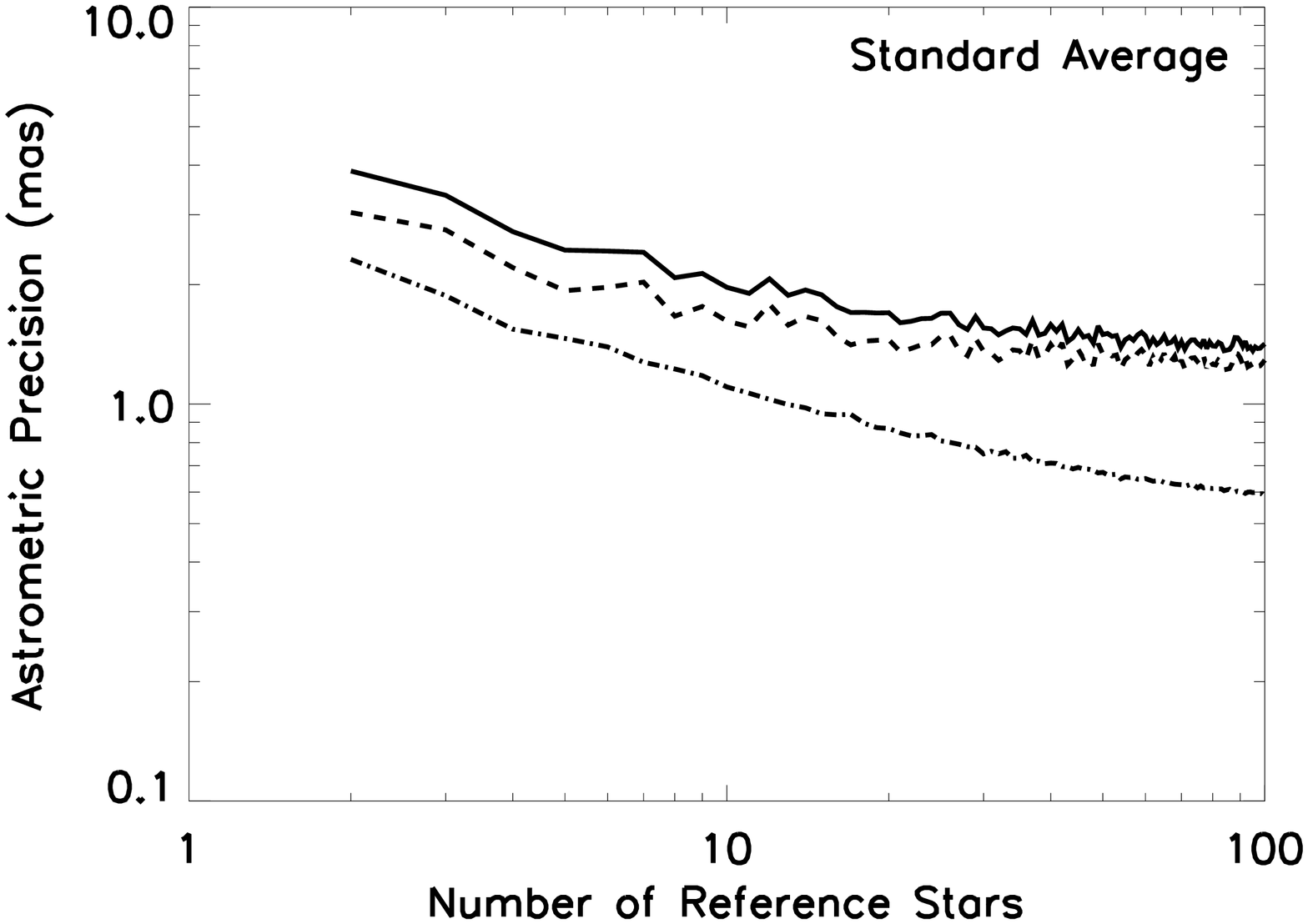}
\includegraphics[angle=90,width=4in]{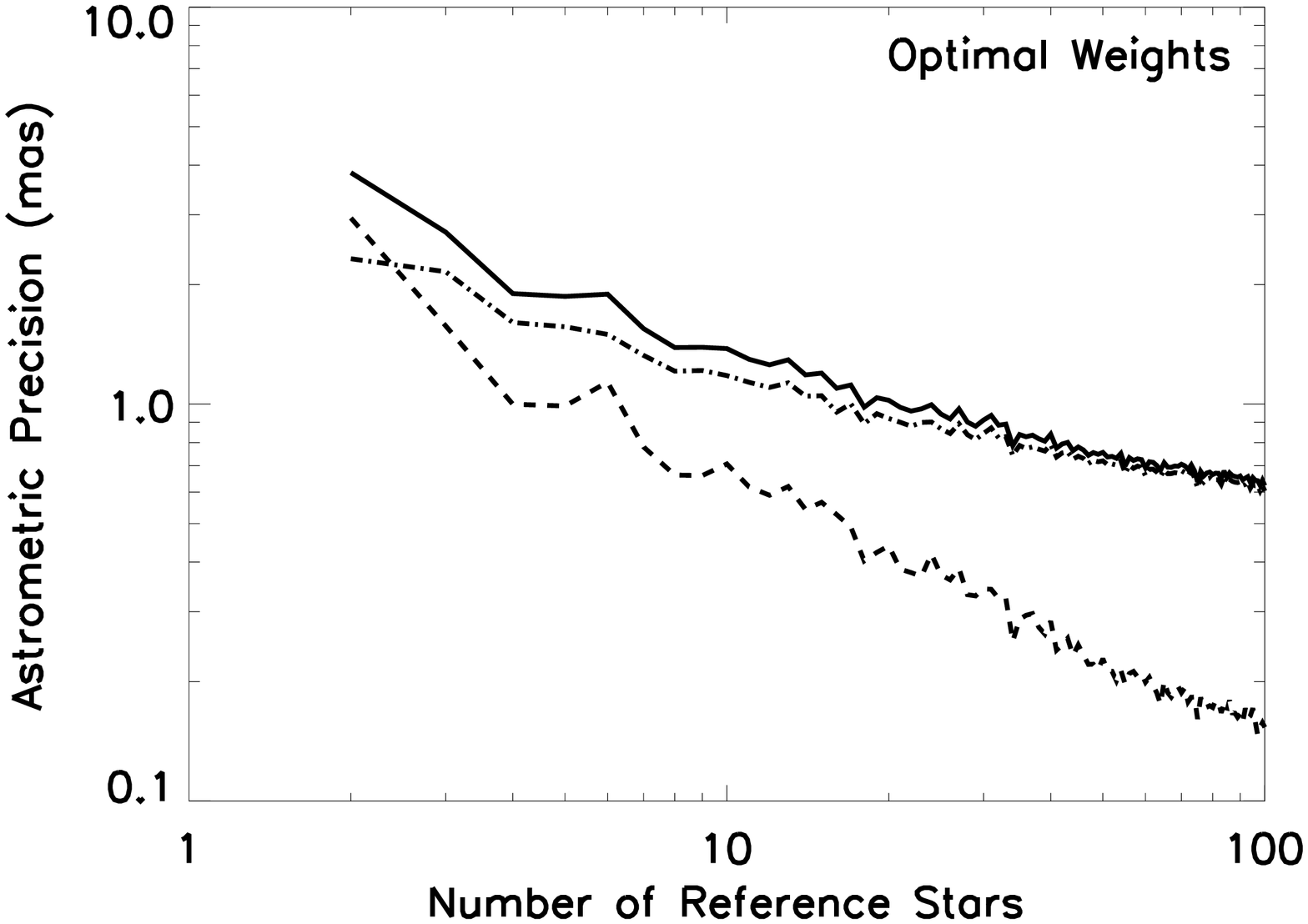}
\end{center}
\caption{{\it Top:} Simulated astrometric precision as a function of
the number of reference stars using standard averaging (solid line). The
total astrometric precision has contributions from the measurement
noise (dash-dotted) and the differential atmospheric tilt jitter
(dashed line). Here the measurement noise was taken to be 0.5\,mas for
the AO guide star, and the reference stars were drawn from a normal
distribution with a mean of 2\,mas and standard deviation of 1\,mas.
The tilt jitter is that expected in a 1.4\,sec exposure assuming the
turbulence profile measured on the night of 2007 May 28 at Palomar
Observatory (see Table~\ref{tab:obs}).  {\it Bottom:} Simulated
astrometric precision as a function of the number of reference stars
using optimal weighting (lines as above).  By using optimal weighting
based on the covariance matrix, the effect of atmospheric noise is
reduced to values less than measurement noise.}
\label{fig:simn}
\end{figure}

The resulting values for the single epoch measurement precision of
$\vec{p}$ are shown in Figure~\ref{fig:simn} as a function of the
number of reference stars, along with the contributions of measurement
noise and differential tilt jitter. In both simulations the error due
to measurement noise decreases as $N^{-0.3}$. However, in the
limit of an infinite number of reference stars, this error asymptotes
to the target star's measurement error. The rate at which the
measurement noise decreases to this value depends on the distribution
of reference star measurement errors.

The important distinction between the two simulations is the
contribution of tilt jitter to astrometric performance. In the
simulation utilizing standard averaging, there is very little gain with
increased stellar density ($N^{-0.15}$), and tilt jitter dominates the
error budget. However, the optimal estimation algorithm rapidly
($N^{-0.7}$) eliminates the contribution of differential tilt by
taking advantage of the correlations inherent in $\mbf[\Sigma_d]$ and
the flexibility to symmetrize the reference field through the choice
of weights. 

\section{Analysis and Results}
\label{sec:res}
In the analysis that follows we will use the measurement model
described in \S\ref{sec:grid}. For a given target star, we will
calculate the differential offsets with respect to the grid stars to
generate a value of $\mbf[d]$ for each image at each epoch
(Equation~\ref{eqn:d}). We then use either these data or the theory in
\S\ref{sec:cvm} to generate the full covariance matrix,
$\mbf[\Sigma_d]$. From $\mbf[\Sigma_d]$, we use the prescription in
\S\ref{sec:optimal} to calculate the the optimal weights,
$\mbf[W]$. These weights are used to combine the differential offsets
to generate the target star's position, $\vec{p}$, in each image via
Equation~\ref{eqn:wavg}. The statistics of the positional measurements
are then described by the covariance matrix, $\mbf[\Sigma]_p$, from
Equation~\ref{eqn:sigp}.

\subsection{Differential Tilt Jitter}

\begin{figure}
%\epsscale{0.65}
\begin{center}
\includegraphics[angle=90,width=4in]{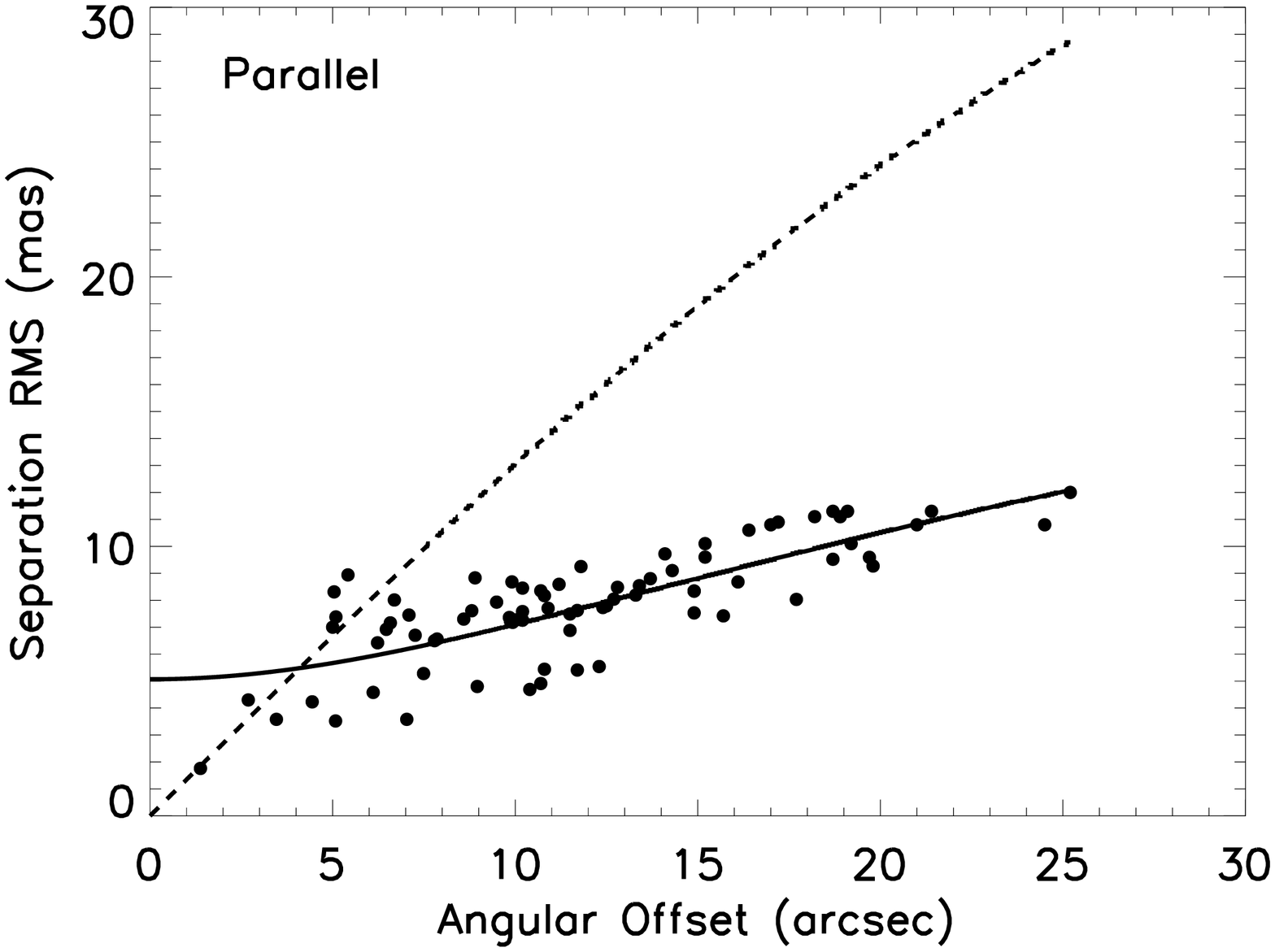}
\includegraphics[angle=90,width=4in]{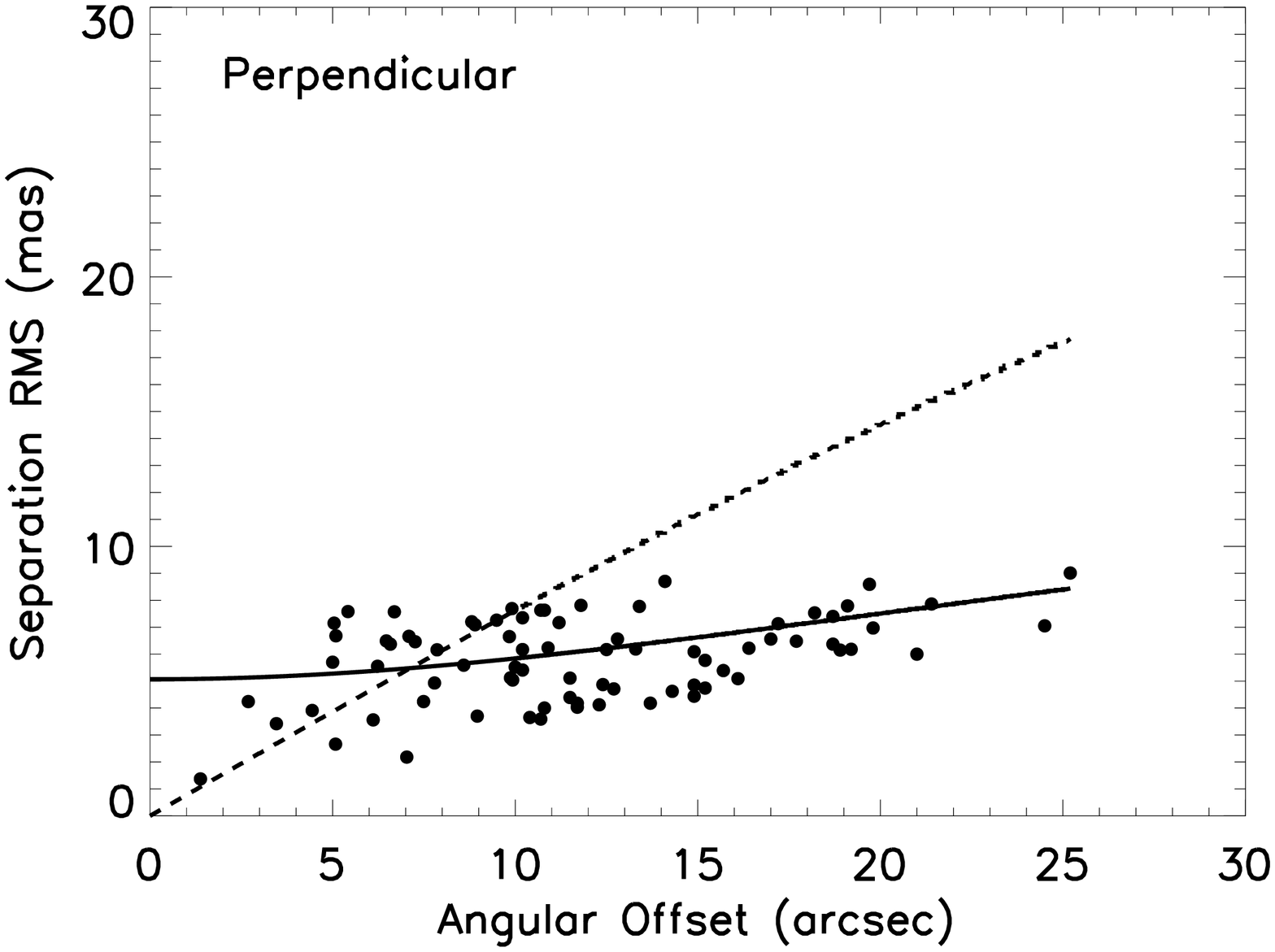}
\end{center}
%\plotone{f3a.ps}
%\plotone{f3b.ps}
\caption{{\it Top:} RMS deviation in the distance between pairs of
stars in the direction parallel to their separation axis on 2007 May
28 (filled circles).  The jitter predicted (assuming no measurement
noise) from the measured turbulence profiles and Equation~\ref{eqn:tj}
(dashed-line) is far larger than the measured jitter, indicating that
some tilt jitter has been averaged away in 1.4\,sec. The best fit
model (Equation~\ref{eqn:varpair}) including averaged tilt jitter and
measurement noise indicates that the tilt timescale is $\approx
0.2$\,sec (solid line).  {\it Bottom:} As above, but in this case the
separations and predictions are for the direction perpendicular to the
separation axis. The expected RMS for the perpendicular direction is
lower by the expected factor as seen in Equation~\ref{eqn:tj}. Note
that not all pairs include the AO guide star.}
\label{fig:tj}
\end{figure}

In order to test our expectation that tilt jitter dominates the
astrometric error, we calculate the RMS of the angular offsets for
pairs of stars in the field (Figures~\ref{fig:finder} and
\ref{fig:tj}).  These results clearly show the characteristic
signature of differential tilt. Namely, the RMS separation along the
axis connecting the two stars is larger than that of the perpendicular
axis by a factor of $\approx \sqrt{3}$.  However, the magnitude of the
tilt jitter is smaller than the theoretical expectations, which
suggests that some of the tilt jitter has been averaged away in the
1.4\,sec exposure time. 

We have no direct measurement of the wind speed profile over the
telescope to calculate the expected tilt jitter timescale.  Instead,
we fit the observed $\sigma_{ij}^2$ and angular offsets using the
model in Equation~\ref{eqn:varpair} with $t=1.4$\,seconds.  The best
fit values are $\sigma_{\rm meas,ij} \approx 2$\,mas and $t/\tau_{\rm
TJ} \approx 7$. This implies that the characteristic timescale for
tilt jitter is $\approx 0.2$\,seconds, resulting in a wind crossing
time of 25\,m\,sec$^{-1}$.  Turbulence at higher altitudes contributes
most to the differential atmospheric tilt jitter, and this velocity is
typical of wind speeds in the upper atmosphere \citep{g77}. It is also
clear from the figure that a number of stars have measurement noise
that is much less than 2\,mas, thus this number should only be taken
as characteristic of the faint stars.

\begin{figure}
\plotone{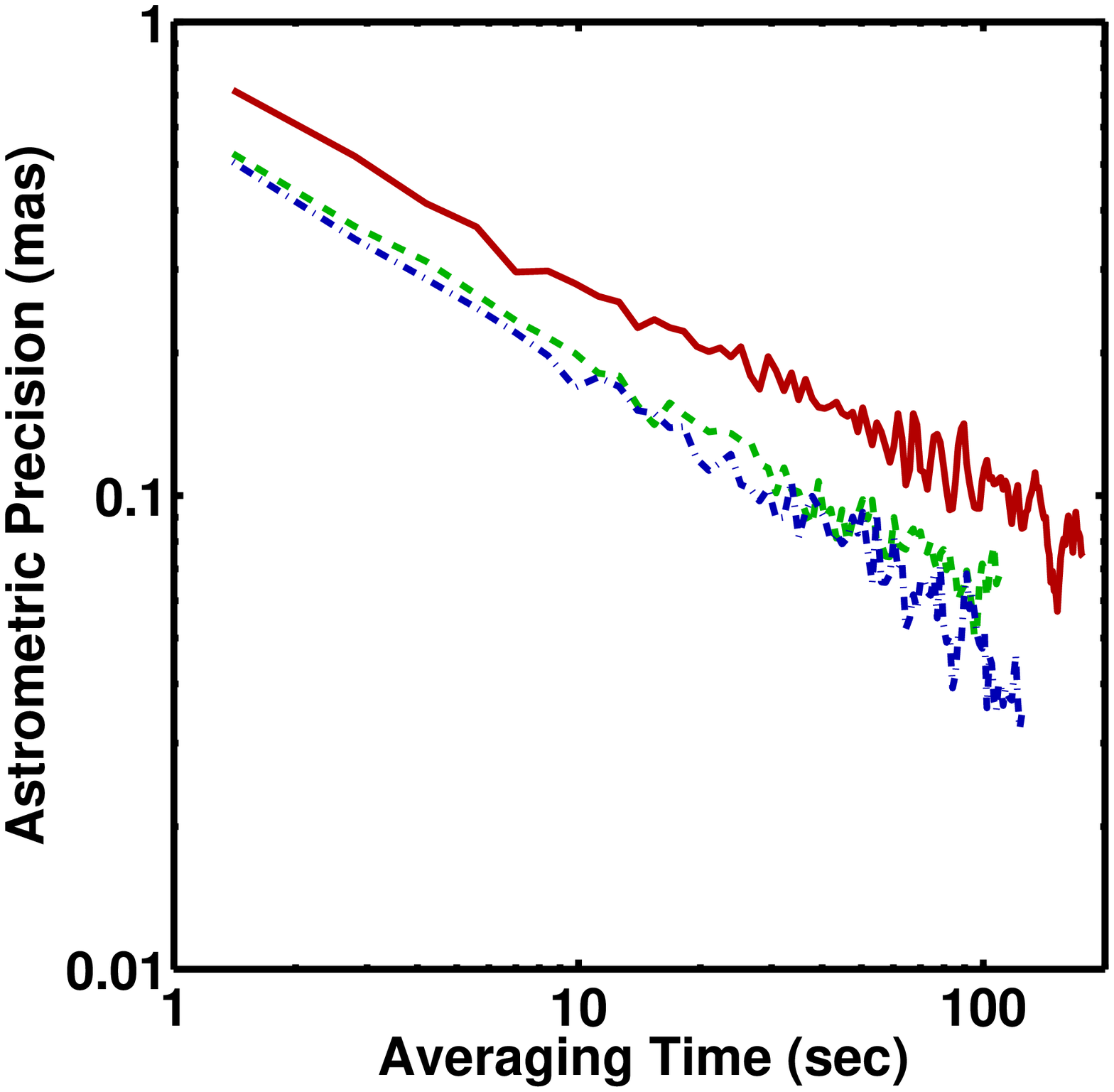}
\caption{Allan deviation in the guide star position as a function of
integration time for 2007 May 28 (solid curve), May 29 (dash-dotted)
and July 22 (dashed). The astrometric precision scales as $t^{-0.51
\pm 0.08}$, and the covariance matrix and optimal weights were derived
from data.}
\label{fig:adevt}
\end{figure}

\subsection{Astrometric Precision}
\label{sec:prec}
The astrometric precision achieved in a single epoch is an important
diagnostic of the measurement model. On a given night for a given
star, we investigate the use of both the $\approx 500$ images and the
theory in \S\ref{sec:cvm} to calculate $\mbf[\Sigma_d]$, leading to
the optimal weights. We then apply this weight matrix to the measured
offsets to compute the target's position in each image --- resulting
in a timeseries in each component of $\vec{p}$ for each epoch.  The
properties of each timeseries are best explored by computing its Allan
deviation (also known as the square root of the two-sample variance).
The Allan deviation is calculated by dividing a timeseries into
chunks, averaging each segment, and computing the RMS of the
resulting, shorter timeseries.  If the timeseries is dominated by
random errors, its Allan deviation will decrease as $1/\sqrt{t_{\rm
avg}}$, where $t_{\rm avg}$ is the length of each chunk. It is also
necessary to have sufficiently many segments so that an RMS
calculation is meaningful. Here the longest timescale probed is
$\approx 2$--3 minutes for each 10--15\,minute timeseries.

We compute the geometric mean of the Allan deviation in each dimension
as a function of the averaging time for the AO guide star in
Figure~\ref{fig:adevt} after computing the covariance matrix from
data.  After 1.4 seconds the guide star's positional precision is
$\approx 600\,\mu$as. The precision subsequently improves as $t^{-0.51
\pm 0.08}$ to $\approx 70\,\mu$as after 2 minutes, and has yet to hit
a systematic floor. This suggests a precision of $\approx 30\,\mu$as
for the full 10--15 minutes data set, assuming that no systematic
limit is reached in the interim.

\begin{figure}
\includegraphics[angle=90,width=6in]{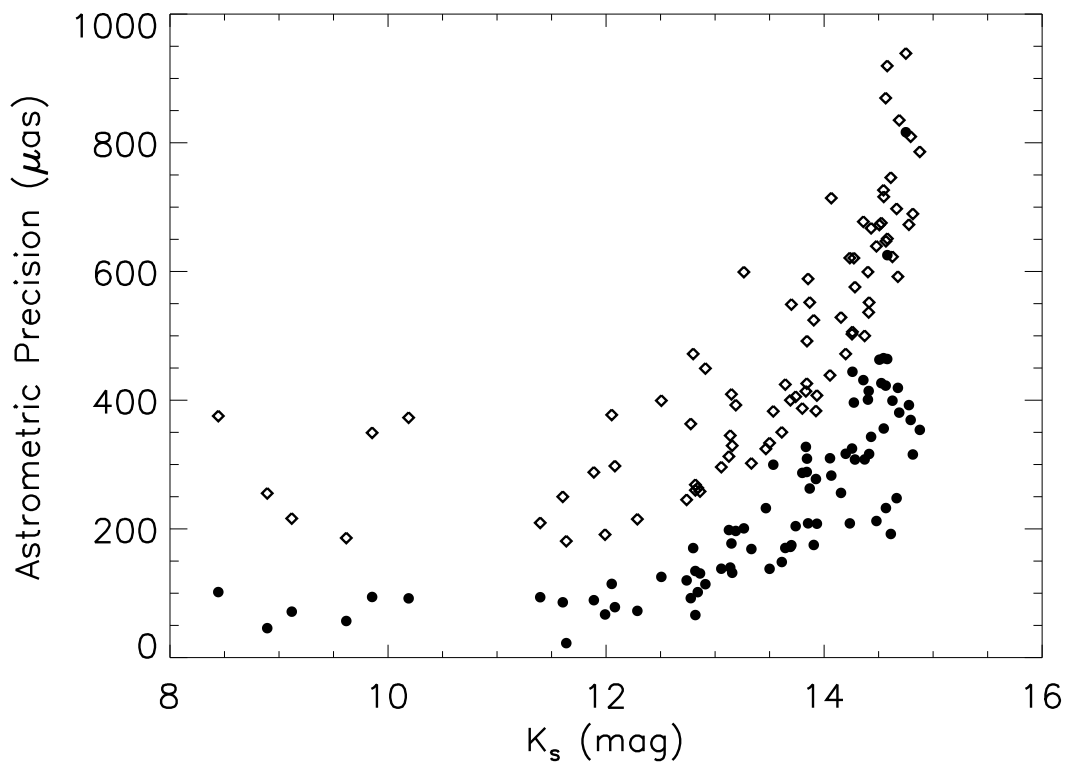} 
\caption{The astrometric precision (Allan deviation after 2 minutes)
using the theoretical covariance matrix (open diamonds) and the
covariance matrix from data (filled circles) as function of $K_s$
magnitude for all 82 detected stars on 2007 May 29. The precision in
both cases is essentially constant for $K_s \simlt 13$\,mag. However,
the astrometric precision for the theoretical $\mbf[\Sigma_d]$ is a
factor of 2--4 times larger than when calculated from data.}
\label{fig:precvmag}
\end{figure}

This level of precision is not limited to the AO guide star; similar
performance is obtained on other stars in the core of M5.  In
Figure~\ref{fig:precvmag} we show the astrometric precision obtained
on 2007 May 29 after 2 minutes for all detected stars as a function of
their $K_s$ magnitude.  Precision below 100\,$\mu$as is achieved on
targets as faint as $K_s \approx 13$ magnitude using a narrow-band
filter and 1.4\,sec individual exposures. This demonstrates the
substantial signal-to-noise ratio benefit afforded by adaptive
optics.

The astrometric precision shown in Figure~\ref{fig:precvmag} resulting
from the theoretically determined covariance matrix and optimal
weights is $\approx 300\,\mu$as after 2 minutes for stars with $K_s
\simlt 13$ magnitude. This level of precision is substantially better
than the performance of simpler weighting schemes, but it is a factor
of 2--4 worse than using the data to calculate the covariance matrix
and weighting. There are several possible reasons for this reduction
in precision. The first is that we have only used estimates of the
measurement noise for each star used to calculate $\mbf[\Sigma]_{\rm
meas}$. Secondly, the turbulence profile used to construct
$\mbf[\Sigma]_{\rm TJ}$ is estimated from the average $C_n^2(h)$ seen
by the DIMM/MASS. This unit is located 300\,m from the Hale telescope
and uses Polaris to estimate the turbulence profile.  As a
consequence, there could be important differences between the measured
atmospheric turbulence and that encountered by the light from
M5. Finally, we have not attempted to capture the time variability of
the turbulence, having used only the average values.

\begin{figure}
\plotone{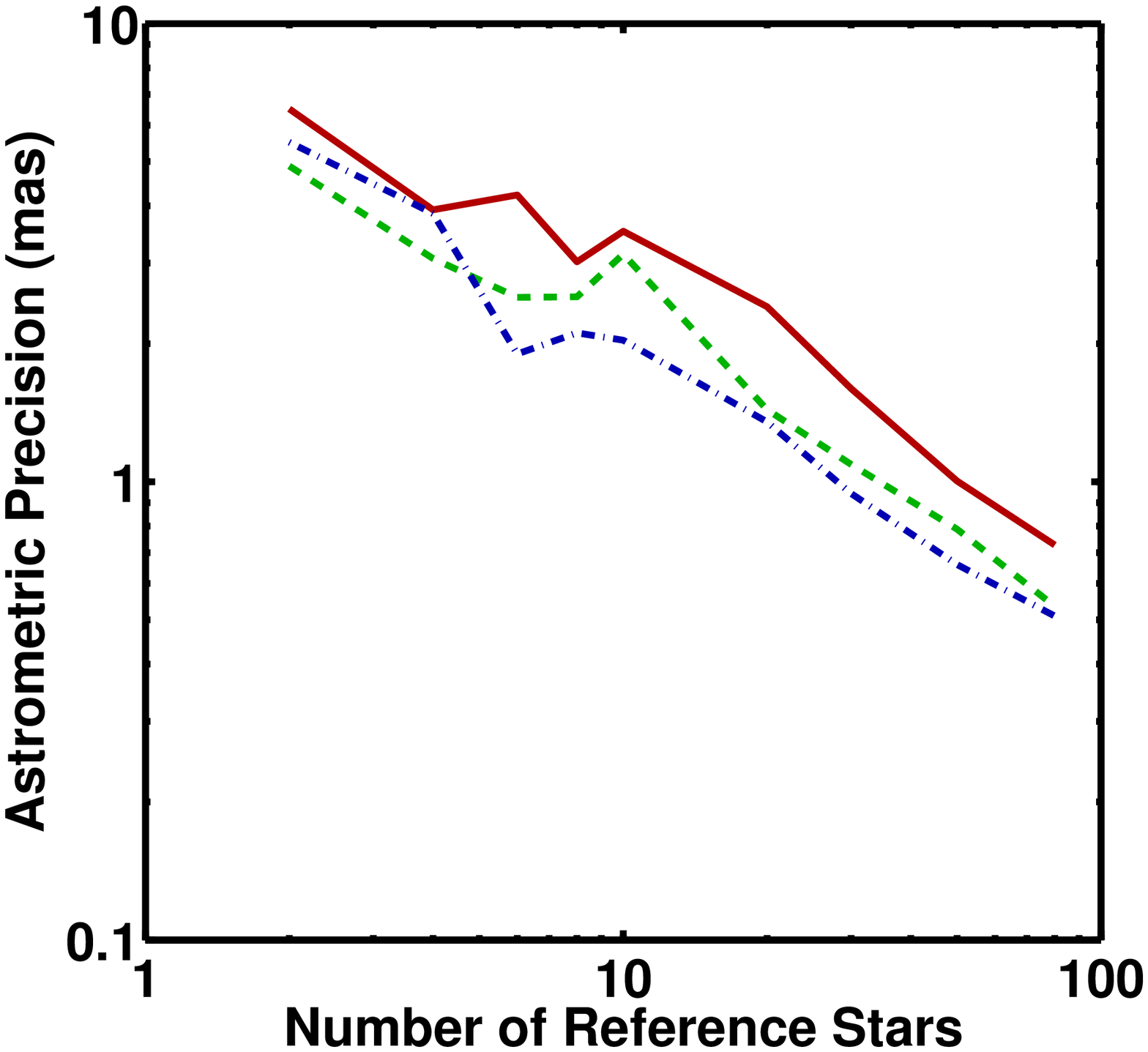}
\caption{Allan deviation in 1.4\,seconds for the AO guide star's
position as a function of the number of reference stars on 2007 May 28
(solid curve), May 29 (dash-dotted), and July 22 (dashed). The
astrometric precision scales as $N^{-0.60 \pm 0.03}$.}
\label{fig:adevn}
\end{figure}

In Figure~\ref{fig:adevn} we investigate the improvement of the AO
guide star astrometry with the number of reference stars. We drew
random subsets of the available grid stars, computed $\mbf[\Sigma_d]$
from the data, calculated the optimal weights, and show the geometric
mean of the eigenvalues of $\mbf[\Sigma_p]$. To average over the
geometry of a particular draw, we repeated this process 10 times for
each value of $N$ and averaged the results. We see that the precision
rapidly decreases as $N^{-0.60 \pm 0.03}$. This is slightly faster
than our simulations predict for 1.4\,sec of integration
time. However, as noted above, our simulations are meant to
approximate M5, but do not capture the true distribution of stellar
measurement errors (which are difficult to decouple from tilt jitter)
or any evolution in atmospheric turbulence during the observation.

\subsection{Astrometric Accuracy}
\label{sec:acc}
The goal of astrometry is to measure the position of the target star
over many epochs. Astrometrically interesting timescales range from
hours to years.  Clearly, the optical systems must be stable over
these spans for astrometry with AO to be viable.  There are several
obstacles that could render the single-epoch precision obtained in
\S\ref{sec:prec} meaningless.  For example, PHARO is mounted at the
Cassegrain focus which results in flexure of the instrument as the
telescope tracks, and undergoes warming and cooling cycles between
observing periods (typically twice per month) that could cause small
changes in the powered optics.  Either of these facts could alter the
geometric distortion, and make astrometric measurements
unrepeatable. In order to probe the system stability, we have designed
our experiment to be as consistent as possible, and it has spanned many
removal and reinstallations of PHARO over 2 months.

\begin{figure}
\plotone{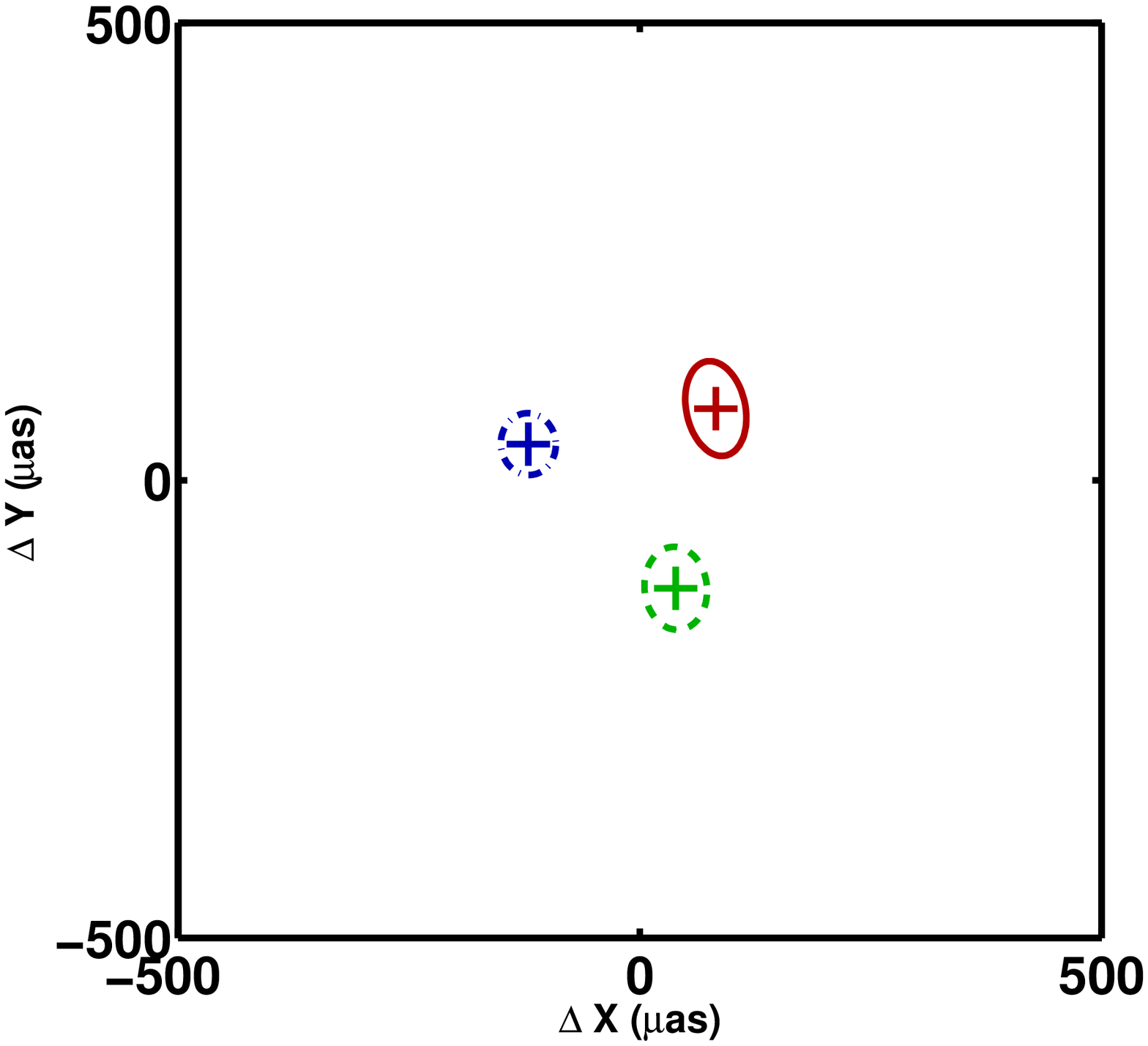}
\caption{The position of the AO guide star in an arbitrary coordinate
system on three dates: 2007 May 28 (solid), 2007 May 29 (dashed), and
2007 July 22 (dash-dotted; see \S\ref{sec:acc}). The error circles are
inferred by averaging the covariance matrix measured from the data and
extrapolating to the total 10--15 minute integration time as
$1/\sqrt{t}$ (e.g. see Figure~\ref{fig:adevt}). The positions agree at
the $\simlt 100\,\mu$as level --- a factor of 2--3 larger than the
expected dispersion. This discrepancy indicates that some systematic
errors have occurred between epochs, most likely optical distortion.}
\label{fig:epochs}
\end{figure}

In order to investigate the accuracy of the M5 measurements we first
measured and corrected the small rotational ($\simlt 0.04^{\circ}$)
and plate scale ($\simlt 10^{-5}$) changes between the the May 29 and
July 22 data and the May 28 images. We also calculate the optimal
weights for a given star on all three nights, and average them to
create one weighting matrix to use for each epoch. This is not
strictly optimal, since each night has different turbulence conditions
for example, but it ensures that the scenario that $\vec{p}
\rightarrow \vec{p} + \vec{\epsilon}$. We see in
Figure~\ref{fig:epochs} that the measured position of the AO guide
star is accurate from epoch-to-epoch at the $\approx 100\,\mu$as.  The
error ellipses are those estimated by continuing to extrapolate the
precision found in Figure~\ref{fig:adevt} by $1/\sqrt{t}$ to the full
10--15 minute timeseries. This is an impressive level of accuracy, but
unfortunately is a factor of 3 worse than our expectation. It suggests
that there is some instability, likely in the distortion, over the two
months that limit the astrometric accuracy.

\begin{figure}
\epsscale{0.65}
\includegraphics[angle=90,width=6in]{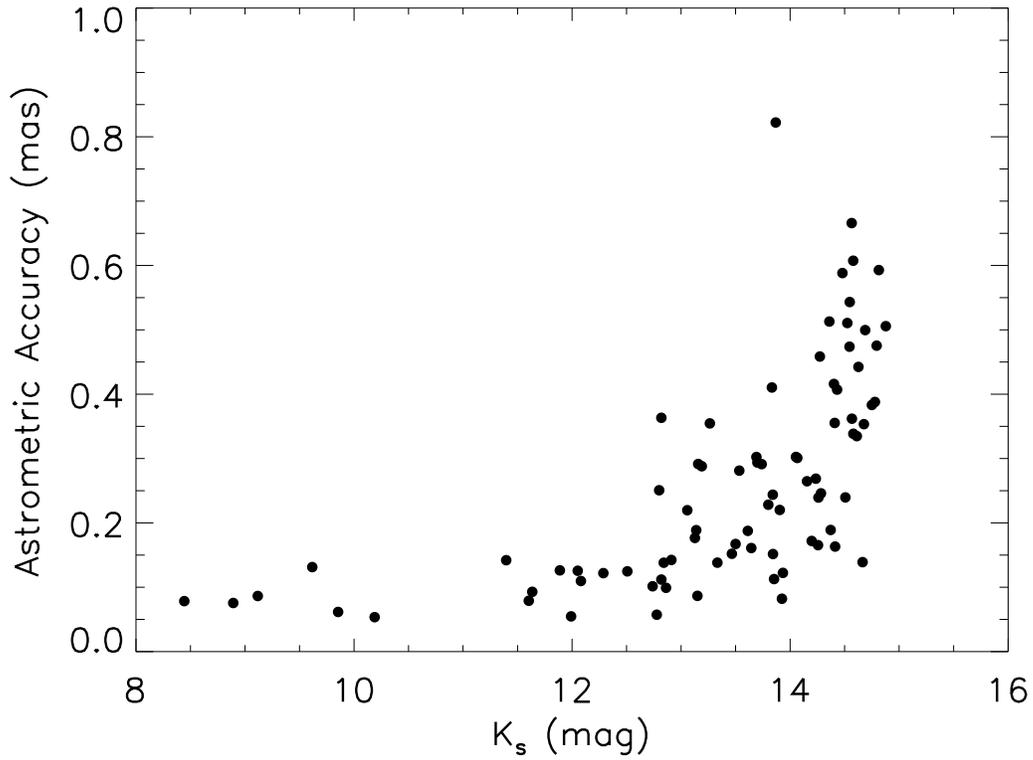}
\caption{Astrometric accuracy (geometric mean of the RMS in each
coordinate over the three epochs) versus $K_s$ magnitude.  The level
of accuracy is $\simlt 100\,\mu$as, and is essentially unchanged for
$K_s \simlt 13$\,mag. However, this is a factor of 2--3 above our
expectations from the achieved astrometric precision, suggesting a
systematic limitation between epochs.}
\label{fig:accuracy}
\end{figure}

The other stars in M5 show a similar level of astrometric accuracy
(Figure~\ref{fig:accuracy}) up to $K_s \approx 13$\,mag.  This limit
can certainly be pushed considerably fainter with in increased
integration time or a larger aperture. The achievement of such high
levels of astrometric performance on faint targets, given the modest
time investment, short integration time and narrow-band filters,
illustrates the substantial signal-to-noise ratio gain and potential
for astrometry enabled by AO.

\section{Discussion and Conclusions}
\label{sec:dis}
Here we have presented a technique for performing high-precision grid
astrometry using ground-based telescopes equipped with adaptive optics
systems.  With this technique, the effects of distortion and
atmospheric dispersion that give rise to systematic errors are
mitigated through the design of the experiment.  Random errors arising
from differential tilt jitter and measurement noise are minimized
through the use of an optimal estimation scheme that accounts for the
correlated noise statistics through the covariance matrix
$\mbf[\Sigma_d]$.  The experimental results obtained on the Hale 200-inch
Telescope have demonstrated single epoch astrometric precision of
$\simlt 100\,\mu$as in 2 minutes and multi-epoch astrometric accuracy
at the same level.  This level of precision is comparable to that
afforded by ground-based interferometry, and is better than the
precision obtained in seeing-limited programs on single apertures.

\begin{figure}
\plotone{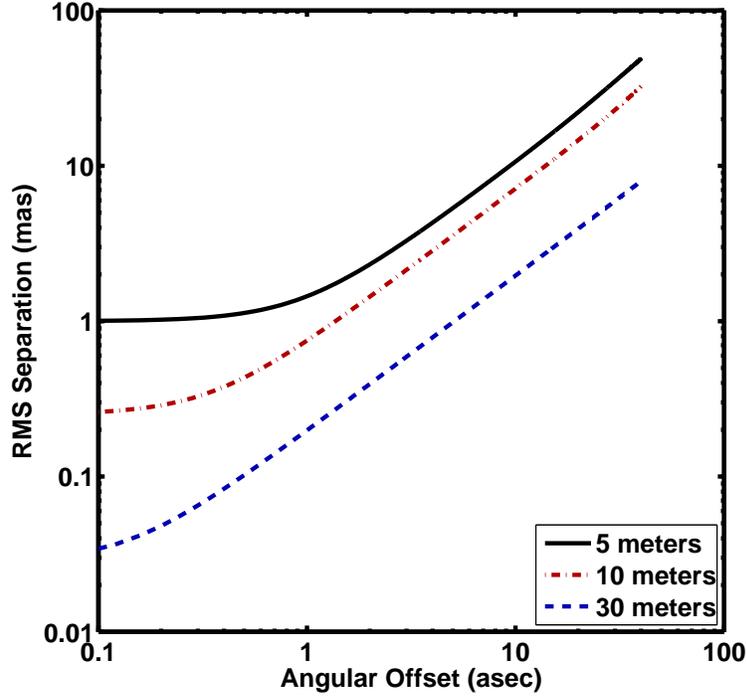}
\caption{The RMS separation between a pair of stars versus angular
offset and aperture diameter. We have assumed the turbulence profile
on 2007 May 28, a measurement error of $1/\sqrt{2}$\,mas for each star
for a 5\,m telescope, and included the geometric mean of each
component of Equation~\ref{eqn:tj}. Relative to Palomar (solid curve),
there are substantial astrometric gains to be made for larger 10\,m
(dash dotted) and 30\,m (dashed) telescopes due to the reduction of
both measurement noise (the $y$-intercept; $\propto D^{-2}$) and tilt
jitter ($\propto D^{-7/6}$).  Because measurement noise falls off more
quickly with $D$, tilt jitter becomes the dominant source of
astrometric error for large aperture telescopes.}
\label{fig:ptj}
\end{figure}

The simulation of astrometric precision afforded by the optimal
weighting scheme shown in Figure~\ref{fig:simn} illustrates that
measurement noise is the dominant residual astrometric error on a 5
meter telescope for stellar fields that contain more than a few
reference stars.  The scaling laws for differential tilt jitter
($D^{-7/6}$) and measurement noise ($D^{-2}$) indicate that on larger
aperture telescopes measurement noise will represent a smaller
fraction of this residual error.  This effect is illustrated in
Figure~\ref{fig:ptj}, which shows the RMS error between pairs of stars
for a range of telescope apertures and angular separations.

The values in Figure~\ref{fig:ptj} assume that tilt jitter is resolved
by sufficiently short exposures. Longer exposure times will certainly
reduce the differential tilt jitter by $1/\sqrt{t}$, but the
measurement noise will also be decreased by this factor (for a given
stellar brightness).  The implication being that if tilt jitter
dominates for short exposure times, it will continue to dominate
longer exposures.

\begin{figure}
\plotone{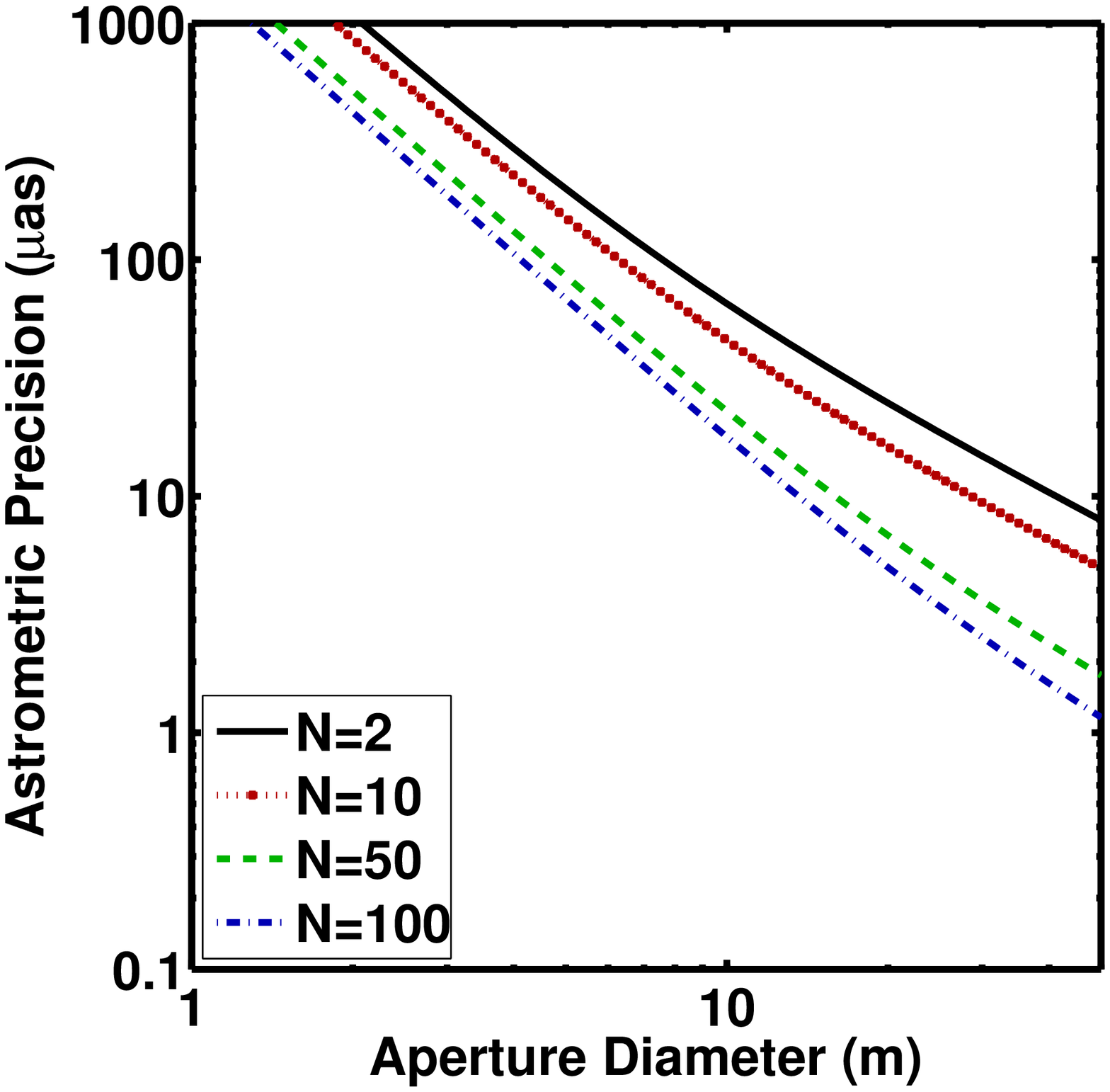}
\caption{Astrometric precision as a function of aperture size and
stellar density. We have used Equation~\ref{eqn:sum} with the
assumptions of the Palomar turbulence profile, a 25\arcsec\ $\times$
25\arcsec\ field of view, the M5 brightness distribution, and photon
noise limit as described in \S\ref{sec:dis}.  The astrometric
precision demonstrates a very favorable scaling law with aperture
diameter, and suggests orders of magnitude improvement in precision
may be available using large aperture, AO equipped telescopes.  In
practice, the level of astrometric accuracy will depend on the extent
to which current and future facilities can characterize and control
systematic errors.}
\label{fig:rmsd}
\end{figure}

In situations where fewer images are available, either due to time
constraints or longer exposure times per frame, it is difficult or
impossible to effectively calculate the covariance matrix directly
from the data. Our results show that independent measurements of the
turbulence profile, for example from a DIMM/MASS unit, are sufficient
to calculate $\mbf[\Sigma_d]$, and result in astrometric precision
within a factor of 2--4 of the levels achieved using the data
itself. Thus, the astrometric applications of turbulence sensors are
two-fold; they can be used to independently assess astrometric data
quality, and predict the AO PSF \citep{b06a}.

The scaling laws presented throughout this paper indicate a
substantially improved astrometric performance on large aperture
telescopes equipped with adaptive optics.  We have used the measured
performance on M5 with the Hale Telescope combined with these scaling
laws to predict the astrometric performance of a single conjugate AO
system as a function of aperture diameter and number of reference
stars.  The relationship can be summarized using the results of
simulation and data analysis as
\be
\sigma_{\rm tot}^2 = \sigma_{\rm meas}^2 + \sigma_{\rm TJ}^2 = 
\left(\frac{1.4{\rm \ sec}}{t}\right)
\left\{
\left[
2{\rm \ mas}\left(\frac{2}{N}\right)^{0.3}\left(\frac{5{\rm \ m}}{D}\right)^2
\right]^2 +
\left[
2{\rm \ mas}\left(\frac{2}{N}\right)^{0.7}\left(\frac{5{\rm \ m}}{D}\right)^{7/6}
\right]^2
\right\}.
\label{eqn:sum}.
\ee
This equation assumes that measurement error is dominated by photon
noise ($\propto D^{-2}$) and the other dependencies (field of view,
stellar brightness distribution, turbulence profile) are identical to
those for the M5 experiment.

Figure 10 shows the resulting estimates for astrometric precision as a
function of aperture diameter and number of reference stars for a 2
minute exposure.  These predictions demonstrate that limits to
astrometric precision arising from random errors (dominated by tilt
jitter) lie below $10\,\mu$as for 30\,m telescopes. However, very
careful characterization and control of systematic errors will be be
required to achieve this level of precision in an actual experiment.
The extent to which systematic errors can be eliminated will
distinguish the scientific goals that can be accomplished with
ground-based facilities from those that require a space-based
solution.

\bigskip\noindent {\it Facilities:} Hale (PALAO/PHARO)

\acknowledgements 

We thank Nicholas Law, Michael Ireland, David Le Mignant, Adam Kraus,
Marten van Kerkwijk, and Andrew Gould for useful discussions on
astrometry. We also thank Palomar Observatory for providing support
for the DIMM/MASS unit used in this study. This work has been
supported by NASA, and by the National Science Foundation Science and
Technology Center for Adaptive Optics, managed by the University of
California at Santa Cruz under cooperative agreement No. AST -
9876783.

\bibliographystyle{apj1b_ams}
\bibliography{ms,journals}

\end{document}